\documentclass[10pt,journal,twocolumns]{IEEEtran}

\usepackage{subfigure}
\usepackage{amssymb}
\usepackage{amsthm}
\usepackage{amsmath}
\usepackage{hyperref}
\usepackage{graphicx}
\usepackage{colortbl,dcolumn}
\usepackage{booktabs}
\usepackage{xcolor}
\usepackage{algorithm}
\usepackage{algorithmic}
\usepackage{cite}

\usepackage{mathtools} 

\newtheorem{definition}{Definition}

\newtheorem{problem}{Problem}
\newtheorem{subproblem}{Subproblem}
\newtheorem{lemma}{Lemma}
\newtheorem{theorem}{Theorem}
\newtheorem{proposition}{Proposition}
\newtheorem{corollary}{Corollary}
\newtheorem{remark}{Remark}

\newcommand{\defref}[1]{Definition~\ref{#1}}

\newcommand{\probref}[1]{Problem~\ref{#1}}
\newcommand{\subpref}[1]{Subproblem~\ref{#1}}
\newcommand{\lemref}[1]{Lemma~\ref{#1}}
\newcommand{\thmref}[1]{Theorem~\ref{#1}}
\newcommand{\propref}[1]{Proposition~\ref{#1}}
\newcommand{\corref}[1]{Corollary~\ref{#1}}
\newcommand{\secref}[1]{Section~\ref{#1}}
\newcommand{\rekref}[1]{Remark~\ref{#1}}
\newcommand{\figref}[1]{Fig.~\ref{#1}}

\newcommand{\algref}[1]{Algorithm~\ref{#1}}
\newcommand{\apxref}[1]{Appendix~\ref{#1}}

\DeclareMathOperator*{\argmin}{argmin}

\graphicspath{{pdfs/},{eps/},{authorphotos/}}

\begin{document}

\title{Event-Trigger Based Robust-Optimal Control for Energy Harvesting Transmitter}

\author{Yirui~Cong,~\IEEEmembership{Student~Member,~IEEE,}
        and~Xiangyun~Zhou,~\IEEEmembership{Member,~IEEE}
\IEEEcompsocitemizethanks{\IEEEcompsocthanksitem
Y. Cong and X. Zhou are with the Research School
of Engineering, Australian National University, Australia (Email: \{yirui.cong, xiangyun.zhou\}@anu.edu.au).\protect\\
}
}

\IEEEtitleabstractindextext{%
\begin{abstract}
This paper studies an online algorithm for an energy harvesting transmitter, where the transmission (completion) time is considered as the system performance. Unlike the existing online algorithms which more or less require the knowledge on the future behavior of the energy-harvesting rate, we consider a practical but significantly more challenging scenario where the energy-harvesting rate is assumed to be totally unknown. Our design is formulated as a robust-optimal control problem which aims to optimize the worst-case performance. The transmit power is designed only based on the current battery energy level and the data queue length directly monitored by the transmitter itself. Specifically, we apply an event-trigger approach in which the transmitter continuously monitors the battery energy and triggers an event when a significant change occurs. Once an event is triggered, the transmit power is updated according to the solution to the robust-optimal control problem, which is given in a simple analytic form. We present numerical results on the transmission time achieved by the proposed design and demonstrate its robust-optimality.
\end{abstract}

}

\maketitle

\IEEEdisplaynontitleabstractindextext

\IEEEpeerreviewmaketitle

\section{Introduction}

\subsection{Motivation and Related Work}

The development of energy harvesting devices has attracted significant attention in recent years with many potential applications in communication networks for green and self-sustainable communications~\cite{GunduzD2014}.
In order to make efficient use of harvested energy, both offline and online solutions have been investigated for designing the optimal transmission policy.
Offline solutions are possible in highly predictable environments where the energy and data arrivals in a sufficiently distant future (for communication purpose) can be accurately estimated~\cite{HoC2013,YangJ2012TWC,YangJ2012TC,TutuncuogluK2012TWC,TutuncuogluK2012JCN,ZaferM2009}.
On the other hand, online solutions typically reduce the dependence on the future knowledge of the energy and data arrival processes, and hence are more applicable in practice.
The online algorithms can be roughly categorized into two frameworks as follows:

In the first framework, the statistical parameters (e.g., the expectation) of energy and data arrival processes are known, and the designs of online transmission policies for energy harvesting nodes are often stated as stochastic control problems.
In~\cite{SharmaV2010,SrivastavaR2013}, the energy and data arrival processes were modeled as stationary and ergodic stochastic processes, where the throughput-optimal and delay-optimal transmission policies were studied.
%
%
By modeling the energy arrival process as a compound Poisson process,~\cite{OzelO2011} proposed a throughput-optimal transmission policy with a deadline in the continuous-time domain by dynamic programming.
%
%
In~\cite{HoC2012}, the energy arrival process was formulated as first-order stationary Markov model and the finite-horizon throughput-optimal transmission policy was derived.
Aiming at minimizing the delay,~\cite{ZhangF2014} provided a closed-form design for the transmission policy which has a multi-level water-filling structure.
Without explicitly modeling the energy and data arrival processes,~\cite{MaoZ2012} employed an upper bound on the long-term data loss ratio and a threshold on the frequency of visits to zero battery state to give a near throughput-optimal transmission policy.
Although the majority of the designed transmission policies aim at either maximizing the throughput or minimizing the delay, other studies also considered maximizing the communication reliability~\cite{WangZ2012} or minimizing the energy consumption~\cite{MaoY2014}.

The second framework uses parameter-independent methodologies.
In~\cite{BlascoP2013}, the energy and data arrival processes were formulated as time-homogeneous Markov chains without knowing the transition matrix, and Q-learning was applied to perform online optimization on the transmission policy.
By Lyapunov optimization technique combined with the idea of weight perturbation,~\cite{HuangL2013_TN} proposed a generic utility-maximization policy, under the assumption that the amount of harvested energy in each time slot is independent and identically distributed (i.i.d.) but its statistical parameters are totally unknown.
Although these parameter-independent policies require less knowledge on the energy and data arrival processes, the stochastic models of the energy and data arrival processes still need to be known exactly.

In practical scenario, the factors determining the energy arrivals are complex, dynamically changing and often unknown to the system designer.
It is sometimes even difficult to come up with accurate models for the energy-harvesting rate.
This leads to an interesting and practical design problem: \textit{how to design and implement an online transmission policy for energy harvesting nodes without imposing any assumption on the future behaviors of the energy-harvesting rate?}
In this paper, we aim to provide an answer to this important question.

Specifically, two advanced methods in cybernetics are employed in this work.
One is the robust optimal control~\cite{ZhouK1996_BOOK}, whose solutions are largely immune to the system uncertainties.\footnote{The term ``robust optimality'' originates from the robust optimization (e.g., in~\cite{BenTalA2009}).
It refers to the optimization of an objective over an uncertain set of situations such that the objective is always not worse than the optimized level.
If all the variables in one robust optimization problem are functions of time, then this problem becomes a robust-optimal control problem.}
We use it to ensure that the system performance (i.e., the transmission time) is no worse than a level (the optimized worst-case performance), no matter what kind of energy arrival process is imposed.
The other method is the event-trigger based control~\cite{AntunesD2014} (or aperiodic control).
It can significantly reduce the unnecessary computations compared to the traditional periodic control (i.e., time-slotted control)\footnote{When designing communication protocol for energy harvesting transmitters, the transmission policy is designed according to the amount of energy and data available.
The event-trigger based control updates the transmission policy only when there is a notable change in the amount of energy.
In contrast, the traditional time-slotted control always performs computation to update the transmission policy at regular time intervals.}.

Our paper is mostly related to the recent work in~\cite{VazeR2014,GomezVilardeboJ2014,VazeR2013}, which considered the similar assumptions that the future energy arrival is unknown.
In~\cite{VazeR2014,GomezVilardeboJ2014,VazeR2013}, the competitive analysis~\cite{BorodinA2005BOOK} was employed to minimize the gap between online and offline performances.
However, minimizing this gap cannot directly guarantee a certain system performance (e.g., optimal worst-case performance) of online algorithms.
Additionally, these recent studies still considered time-slotted systems, and hence, the transmission protocol is updated in every time slot, regardless of the change in the amount of energy available.

\subsection{Our Contributions}

In this work, we study the performance of an energy-harvesting transmitter measured by the transmission (completion) time, i.e., the time duration it takes to complete the transmission of a given amount of data.
We propose to use event-trigger based design to control the transmit power without any knowledge on the future behavior of the energy-harvesting rate.
In the considered scenario, it is not possible to use any statistics of the transmission time in the design.
Hence, we adopt the robust-optimal control to minimize the worst-case transmission time.
Nevertheless, the minimum worst-case transmission time may not always be finite.
When the minimum worst-case transmission time is infinite, which happens when too much data is given to be transmitted with insufficient initial battery energy, we measure the robust-optimality of the transmit power design by looking at the set of energy-harvesting rates that result in finite transmission times.
The robust-optimal design ensures the largest set of energy-harvesting rates resulting in finite transmission times.

The proposed event-trigger based transmitter has two building blocks for implementing the transmit power, namely an Event Detector (ED) and a Transmission Planner (TP).
The ED continuously monitors the battery energy and triggers a new event when it experiences some significant change since the last event.
Whenever an event is triggered, the TP uses the current knowledge of battery energy and data queue to update the transmit power by robust-optimal control.
The updated transmit power is implemented until the next event is triggered.
To the best of our knowledge, this is the first time that the event-trigger based design is implemented on energy harvesting transmitters.

To facilitate the robust-optimal design, we first give a comprehensive analysis on the behavior of battery energy and data queue in each triggered event.
Specifically, we define the reachable set, which describes all possible states (battery energy and data queue) reachable in one event based on the TP's knowledge, and reflects the relationship among battery energy, data queue and transmit power.
Base on these analyses, we derive the solution of the robust-optimal transmit power design, given in a simple analytic form.

\subsection{Paper Organization and Notation}

In \secref{sec:System Model and Event-Trigger Method}, the system model is given and the event-trigger based transmission is introduced.
In \secref{sec:Problem Description}, the problem of finding the robust-optimal transmit power design is defined.
We study the properties of the proposed event-trigger based system through the reachable set analysis in \secref{sec:Reachable Set Analysis}.
The optimal solution to the problem is given in \secref{sec:Solution to RTT Problem and Discussion on Triggering Condition}.
In \secref{sec:Simulation Results}, simulation results are shown to illustrate the effectiveness of our design and corroborate our theoretical results.
Finally, conclusion is drawn in \secref{sec:Conclusion}.

Throughout this paper, $\overline{\mathbb{R}}_+$, $\mathbb{R}_+$, and $\mathbb{Z}_+$ denote the sets of non-negative real numbers, positive real numbers, and positive integers.
%
%
%
$\dot{X}(t)$ denotes the time derivative of $X(t)$ which is a function of time.
$\mu(S)$ is the Lebesgue measure of set $S$.
The restriction (Page 36 in~\cite{StollR1979BOOK}) of function $f$ to domain $\mathcal{A}$ is $f|_{\mathcal{A}}$.
For $x \in \mathbb{R}$, $[x]^+$ returns $\max\{x, 0\}$.

\section{System Model and Event-Trigger Approach}\label{sec:System Model and Event-Trigger Method}

\subsection{System Model}\label{sec:System Model}

\begin{figure}
\centering
\includegraphics [width=1.0\columnwidth]{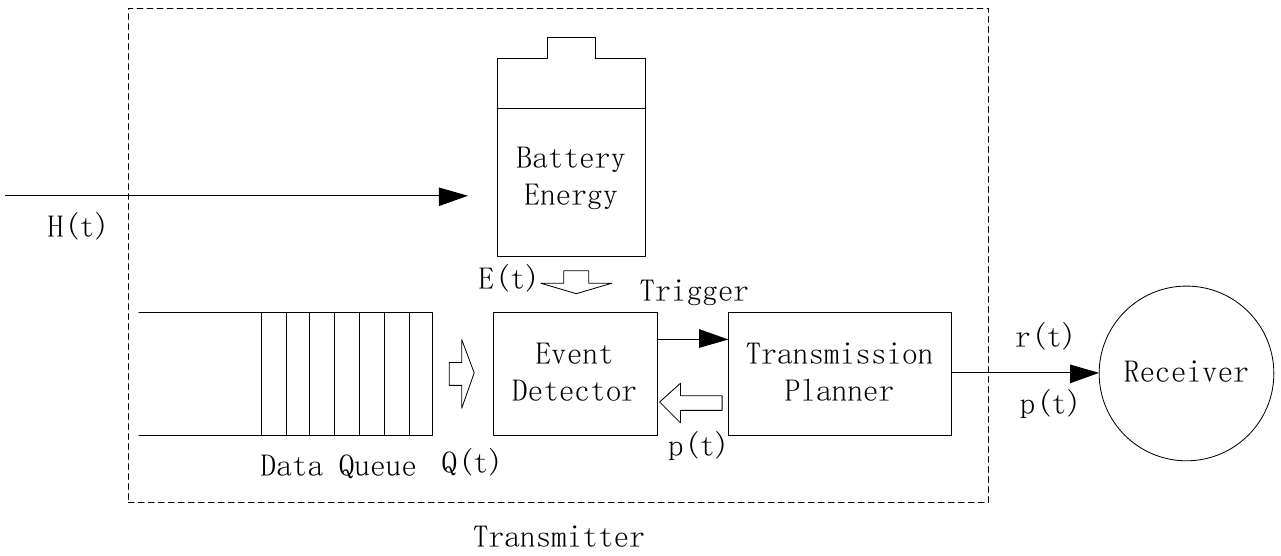}

\caption{The event-trigger based energy harvesting transmitter.}\label{fig:Transmitter and Receiver}
\end{figure}

We consider a transmitter-receiver pair as shown in Fig.~\ref{fig:Transmitter and Receiver}.
At time $t \in [t_0,~\infty)$, where $t_0$ is the starting time of the communication, the transmitter has a battery with energy $E(t) \in \overline{\mathbb{R}}_+$.
The harvested energy is stored in the battery with energy-harvesting rate $H: [t_0,~\infty) \to \overline{\mathbb{R}}_+$.
We assume $H$, as a function of time, is Lebesgue integrable over any subset of $\overline{\mathbb{R}}_+$ with finite measure, and all such $H$ form the set $\mathcal{H}$.
The transmit power at time instant $t$ is $p(t) \in [0,~p_{\max}]$, which is also Lebesgue integrable over any subset of $\overline{\mathbb{R}}_+$ with finite measure, and $p_{\max}$ denotes the maximum power constraint.
Then, the relationship among battery energy, energy-harvesting rate and transmit power is given by a differential equation
\begin{align}\label{eqn:Battery Energy Equation}
\dot{E}(t) = H(t) - p(t),
\end{align}
where the initial battery energy is $E(t_0) \in \overline{\mathbb{R}}_+$.
In this paper, we consider the transmission-time minimization problem, where all the data to be transmitted is available at $t_0$.
The data queue is $Q(t) \in \overline{\mathbb{R}}_+$, and the transmission rate at $t$ is $r(t) \in \overline{\mathbb{R}}_+$, which is Lebesgue integrable over any subset of $\overline{\mathbb{R}}_+$ with finite measure, and the relationship between data queue length and transmission rate is
\begin{align}\label{eqn:Data Queue Equation}
\dot{Q}(t) = - r(t),
\end{align}
where the initial data queue is $Q(t_0) \in \mathbb{R}_+$.
Equation~\eqref{eqn:Data Queue Equation} means all the amount of data (equal to $Q(t_0)$) to be transmitted is available at $t_0$, and there is no subsequent data arrival in $(t_0,~\infty)$, which is a commonly used assumption for transmission-time minimization problem (see~\cite{YangJ2012TC} for an example).
The other system assumptions are as follows:
\begin{itemize}
\item   The energy-harvesting rate $H$ is totally unknown.
\item   Battery energy $E(t)$ can be measured at current time $t$.
\item   The channel is assumed to be static such that the channel capacity is $C(t) = \log_2(1 + p(t))$.
\end{itemize}

\begin{remark}
Unlike most existing online algorithms which somewhat included the prior-knowledge on $H$, we aim to derive an online algorithm without any knowledge of $H$.
%
%
Additionally, the battery energy can be monitored continuously.\footnote{The technology for continuously monitoring the battery energy with minimal operating energy consumption has been developed over the past decades (e.g.,~\cite{JohnsonL1987Patent}). Hence, we assume that this operating energy can be neglected as compared to the energy consumption for transmission.}
We consider a simple static channel that is only affected by AWGN. Note that the static-channel assumption is widely used in the literature, e.g.,~\cite{YangJ2012TC,DevillersB2012,TutuncuogluK2012TWC}.
\end{remark}

Note that the transmit power $p(t)$ and the transmission rate $r(t)$ are to be designed by us, and in this paper, we assume the channel capacity is achieved, i.e., $r(t) = C(t) = \log_2(1 + p(t))$, which means at each time $t$ the transmission rate is a function of transmit power, labeled as $r(p(t))$.
Now, the only variable to be designed is the mapping $p: t \mapsto p(t)$.



\subsection{Event-Trigger Based Control}\label{sec:Event-Trigger Based Transmission Policy}

As shown in Fig.~\ref{fig:Transmitter and Receiver}, the event-trigger control relies on an Event Detector (ED) that detects the necessary changes in the cumulated energy; and a Transmission Planner (TP) that gives the design for transmit power $p(t)$.
%

We give the condition under which an event is triggered.

\begin{definition}[Triggering Condition]\label{def:Trigger Condition}
From a given time instant $s$, an event is triggered at $t~(t>s)$ whenever the following inequalities is satisfied
\begin{align}
\int_s^t H(\tau)\mathrm{d}\tau \stackrel {(a)} {=} E(t) - E(s) + \int_s^t p(\tau)\mathrm{d}\tau \geq \varepsilon,\label{eqn:Event E Happens}
\end{align}
where $\varepsilon \in \mathbb{R}_+$ is the triggering threshold.
\end{definition}

In Definition~\ref{def:Trigger Condition}, condition~\eqref{eqn:Event E Happens} means that when the harvested energy cumulates over a certain level $\varepsilon$, the ED triggers a new event.
Even though $H(t)$ is unknown, the integrals of $H(t)$ can be calculated by equality $(a)$ in~\eqref{eqn:Event E Happens}, which is derived by the solutions of~\eqref{eqn:Battery Energy Equation}, where $E(t)$ is observable.
The transmit power $p(\tau)$ for $\tau \in (s,~t)$ is determined by the TP (to be discussed in later part of this subsection).
%
%
%


Recall that the transmission is carried out over the entire communication time interval $[t_0,~\infty)$. At the initial time instant $t_0$, the ED triggers the start of the transmission. Then, the ED will start monitoring the system on $[t_0,~t]$, where $t$ is the current time instant.
We label the first time instant (after $t_0$) at which the system satisfies the triggering condition in \defref{def:Trigger Condition} by $t_1$.
After $t_1$, the ED will start monitoring the system on $(t_1,~t]$. The next time instant at which an event is triggered is labeled as $t_2$, and so on.
For convenience, we say that event $n$ starts at $t_n$ and finishes at $t_{n+1}$.
%
%
This completes our description of the ED.


Whenever an event comes, the TP plans the transmit power to be implemented from the current time instant until the next event arrives, and we analyze event $n$ without loss of generality.
It is important to note that the TP only takes into account the information available at the beginning of the event when planning the transmit power.
Such information includes the battery energy and data queue length at $t_n$, i.e., $E(t_n)$ and $Q(t_n)$.
However, any future change due to energy-harvesting rate, i.e., $H(t)$ for $t > t_n$, cannot be taken into account, simply because $H(t)$ for $t > t_n$ is unknown to the TP at $t_n$.
Specifically, at the beginning of the $n$\textsuperscript{th} event, i.e., at the time instant $t_n$, the TP records the values of $E(t_n)$ and $Q(t_n)$, and designs $p(t)$ to be implemented over $(t_n,\,t_{n+1}]$ using~\eqref{eqn:Battery Energy Equation} and~\eqref{eqn:Data Queue Equation} with $H(t) = 0$.
This does not mean the TP neglects the effect from $H(t)$ all the time, because $H(t)$ determines the arrival time of the next event, by the triggering condition.

In each event $n$, the TP plans the transmit power for a finite time window after $t_n$, and we call it as the Planned Transmit Power (PTP):
\begin{align}\label{eqn:Planned Transmission Policy}
\tilde{p}_{n,\varepsilon}: (t_n,\,t_n+T_n] \to [0,~p_{\max}],
\end{align}
where the subscript $\varepsilon$ means the PTP is designed under a given triggering threshold $\varepsilon$, and the duration of the time window $T_n$ is called as the planned transmission time.
In~\eqref{eqn:Planned Transmission Policy}, the PTP is described as a mapping from a time instant to a value of transmit power.
For any $t \in (t_n,~t_n+T_n]$, $\tilde{p}_{n,\varepsilon}(t)$ is the value of transmit power at time $t$.
%
%
Thus, a PTP can be described by two parameters, i.e., $\tilde{p}_{n,\varepsilon}(t)$ and $T_n$.
For example, the PTP $\tilde{p}_{n,\varepsilon}(t) = |\sin t|,~t \in (t_n,\,t_n+1]$ is described by pair $(\tilde{p}_{n,\varepsilon}(t), T_n) = (|\sin t|, 1)$.
All possible $\tilde{p}_{n,\varepsilon}$ compose the PTP set $\widetilde{\mathfrak{P}}_{n,\varepsilon}$.
The design problem here is to find a good mapping $t \mapsto \tilde{p}_{n,\varepsilon}(t)$ in~\eqref{eqn:Planned Transmission Policy}.
Note that $T_n$ is part of the description of a PTP, and in other words, the design of $\tilde{p}_{n,\varepsilon}$ effectively includes the design of $T_n$.
We will see in \secref{sec:Solution to RTT Problem and Discussion on Triggering Condition} that $T_n$ plays an important role in designing the optimal $\tilde{p}_{n,\varepsilon}$.
%

Ideally, the PTP should be implemented over $(t_n,\,t_n + T_n]$.
However, the TP performs the design at $t_n$ based on the current information ($E(t_n)$ and $Q(t_n)$), and hence, cannot predict the exact value of $t_{n+1}$, i.e., it does not know when the next event occurs.
As a result, the PTP in event $n$ will not be implemented beyond $t_{n+1}$ because a new PTP will be planned and implemented after $t_{n+1}$.
Hence, the actual transmission time is $\min\{T_n, t_{n+1} - t_n\}$.
It implies that when the ($n+1$)\textsuperscript{th} event comes, the TP will use the newly designed PTP for event $n+1$, even if the planned transmission for event $n$ is unfinished.
Therefore, the \textit{actual transmit power}, denoted by $p_{\varepsilon}$ is implemented piecewise by $\left.\tilde{p}_{n,\varepsilon}\right|_{(t_n,~t_{n+1}]}$ (restricting $\tilde{p}_{n,\varepsilon}$ to $(t_n,~t_{n+1}]$) for each event $n$.
Specifically, the relationship between the actual transmit power and the PTP is
\begin{align}\label{eqn:Relationship between Transmit Power and PTP}
p_{\varepsilon}(t) = \tilde{p}_{n,\varepsilon}(t), \quad t \in (t_n,~t_{n+1}].
\end{align}
For a given triggering threshold $\varepsilon$, all such $p_{\varepsilon}: t \mapsto p_{\varepsilon}(t)$ in~\eqref{eqn:Relationship between Transmit Power and PTP} form the set of all possible transmit power $\mathfrak{P}_{\varepsilon}$.
Recall that $p_{\varepsilon}$ is Lebesgue integrable over any subset of $\overline{\mathbb{R}}_+$ with finite measure.
Thus, $\mathfrak{P}_{\varepsilon}$ is a set of non-negative Lebesgue integrable functions over any subset of $\overline{\mathbb{R}}_+$ with finite measure.

To sum up, the event-trigger control framework is illustrated in~\algref{alg:Event-Trigger Based Control}, where Lines~\ref{line:Design PTP 1} and~\ref{line:Design PTP 2} are the very part to be designed in the rest of this paper.

\begin{algorithm}
\begin{footnotesize}
\caption{Event-Trigger Based Control}\label{alg:Event-Trigger Based Control}
\begin{algorithmic}[1]
\STATE  \textbf{Initial Condition:} $t = t_0$, $t_n = t_0$, $E(t) = E(t_0)$, $Q(t) = Q(t_0)$, $E(t_n) = E(t_0)$, and $Q(t_n) = Q(t_0)$.
\STATE  Assuming $H(t) = 0$ for $t > t_n$, the TP design the PTP $\tilde{p}_{n,\varepsilon}$ (see~\eqref{eqn:Planned Transmission Policy});\label{line:Design PTP 1}
\WHILE  {$Q(t) > 0$}
    \STATE  The ED updates $E(t)$ and $Q(t)$ (the update frequency is dependent on the chip's clock) and checks the condition in~\eqref{eqn:Event E Happens} with $s = t_n$;
    \IF {Condition~\eqref{eqn:Event E Happens} is satisfied}
        \STATE  The ED triggers an event to activate the TP;
        \STATE  The TP updates $t_n = t$, $E(t_n) = E(t)$, and $Q(t_n) = Q(t)$;
        \STATE  Assuming $H(t) = 0$ for $t > t_n$, the TP design the PTP $\tilde{p}_{n,\varepsilon}$ for this event;\label{line:Design PTP 2}
        \STATE  The transmitter uses the newly designed PTP as the transmit power, i.e., $p(t) = \tilde{p}_{n,\varepsilon}(t)$;
    \ELSE
        \STATE  TP is inactive;
        \STATE  The transmitter uses the most recent PTP as the transmit power, i.e., $p(t) = \tilde{p}_{n,\varepsilon}(t)$;
    \ENDIF
\ENDWHILE
\end{algorithmic}
\end{footnotesize}
\end{algorithm}


\section{Problem Description}\label{sec:Problem Description}

In this paper, we study the transmission-time minimization problem under unknown energy-harvesting rate $H$.
The transmission time is the time spent by the transmitter to clear up the data queue, and we label it as $\mathcal{T}(p_{\varepsilon}, E(t_0), Q(t_0), H)$, which is dependent on the transmit power design $p_{\varepsilon}: t \mapsto p_{\varepsilon}(t)$, the initial battery energy $E(t_0)$, the initial data queue $Q(t_0)$, and the energy-harvesting rate $H$.

Since $H$ is totally unknown, given $p_{\varepsilon}$, $E(t_0)$, and $Q(t_0)$, the transmission time varies with different $H \in \mathcal{H}$.
As a result, the transmission time is within the following range
\begin{multline}\label{eqn:Transmission Time Range for a Given Transmit Power Design}
\inf_{H \in \mathcal{H}} \mathcal{T}(p_{\varepsilon}, E(t_0), Q(t_0), H) \leq \mathcal{T}(p_{\varepsilon}, E(t_0), Q(t_0), H)\\ \leq \sup_{H \in \mathcal{H}} \mathcal{T}(p_{\varepsilon}, E(t_0), Q(t_0), H).
\end{multline}
Note that it is impossible to compute the average (or other statistical properties of) transmission time for any given design because the statistics of $H$ are totally unknown.
Nevertheless, it is possible to examine the worst-case transmission time for any given design, i.e., $\sup_{H \in \mathcal{H}} \mathcal{T}(p_{\varepsilon}, E(t_0), Q(t_0), H)$.
Thus, our approach is to find a design that achieves the minimum worst-case transmission time.
In other words, our design aims to give the best performance under the worst-case scenario.
%
Based on this idea, the transmission-time-minimization problem is defined in \subpref{subp:Transmission-Time-Minimization Problem}.

\begin{subproblem}[Transmission-Time-Minimization Problem]\label{subp:Transmission-Time-Minimization Problem}
Given initial battery energy $E(t_0)$, initial data queue $Q(t_0)$, and triggering threshold $\varepsilon$, design PTP $\tilde{p}_{n,\varepsilon}$ in each event $n$, with the knowledge of $E(t_n)$ and $Q(t_n)$, such that
\begin{align}\label{eqn:Transmission-Time-Minimization Problem}
\mathcal{T}^* = \inf_{p_{\varepsilon} \in \mathfrak{P}_{\varepsilon}} \sup_{H \in \mathcal{H}} \mathcal{T}(p_{\varepsilon}, E(t_0), Q(t_0), H),
\end{align}
where the actual transmit power $p_{\varepsilon}$ is determined by the PTP $\tilde{p}_{n,\varepsilon}$ as shown in~\eqref{eqn:Relationship between Transmit Power and PTP}.
\end{subproblem}



\begin{remark}
Indeed, the idea of defining \subpref{subp:Transmission-Time-Minimization Problem} is borrowed from robust-optimal control~\cite{ZhouK1996_BOOK}:
In~\eqref{eqn:Transmission-Time-Minimization Problem}, the $\sup$ operator returns the worst transmission time for a given $p_{\varepsilon}$ under its corresponding worst-case energy-harvesting rate $H_{p_{\varepsilon}}$; while the $\inf$ operator reflects our aim of designing $p_{\varepsilon}^*$ whose worst transmission time $\mathcal{T}\left(p_{\varepsilon}^*, E(t_0), Q(t_0), H_{p_{\varepsilon}^*}\right)$ is the smallest.\footnote{The subscript $p_{\varepsilon}$ in $H_{p_{\varepsilon}}$ highlights the fact that the worst-case $H$ depends on the given transmit power. Although, mathematically $H_{p_{\varepsilon}}$ (for any $p_{\varepsilon}$) might not exist in $\mathcal{H}$, since the operator in~\eqref{eqn:Transmission-Time-Minimization Problem} is $\sup$ rather than $\max$, the existence of the optimal transmit power design $p_{\varepsilon}^*$ is given in \thmref{thm:Optimal Solution to RTT Problem}.}
One technical challenge in solving \subpref{subp:Transmission-Time-Minimization Problem} is that the worst-case energy-harvesting rate depends on the choice of transmit power, i.e., for different $p_{\varepsilon}$, the worst-case $H_{p_{\varepsilon}}$ can be different.
\end{remark}

It is important to note that $\mathcal{T}^*$ is not always finite.
In the case $\mathcal{T}^* = \infty$, equation~\eqref{eqn:Transmission-Time-Minimization Problem} can hardly measure the robust optimality on the designed $p_{\varepsilon}$, since for any $p_{\varepsilon}$ the worst-case transmission time is always infinite.
Hence, \subpref{subp:Transmission-Time-Minimization Problem} is not sufficient for describing all scenarios and we need a different problem formulation to deal with the case of $\mathcal{T}^* = \infty$ as explained as follows:
For a given $p_{\varepsilon}$, there should exist some energy-harvesting rate $H$ resulting in a finite transmission time, even though the worst-case $H_{p_{\varepsilon}}$ may lead to an infinite transmission time.
All possible such energy-harvesting rates form the finite-transmission-time energy set $\mathcal{H}_f(p_{\varepsilon},E(t_0),Q(t_0)) \subset \mathcal{H}$, defined in \defref{def:Finite-Transmission-Time Energy Set}.

\begin{definition}[Finite-Transmission-Time Energy Set]\label{def:Finite-Transmission-Time Energy Set}
Given $p_{\varepsilon}$, $E(t_0)$, and $Q(t_0)$, the finite-transmission-time energy set $\mathcal{H}_f(p_{\varepsilon},E(t_0),Q(t_0))$ is
\begin{align}\label{eqn:Finite-Transmission-Time Energy Set}
\left\{H\colon \mathcal{T}(p_{\varepsilon},E(t_0),Q(t_0),H) < \infty, \quad H \in \mathcal{H}\right\}.
\end{align}
\end{definition}

Considering two transmit power designs, denoted by $p_{\varepsilon}^{\mathrm{a}}$ and $p_{\varepsilon}^{\mathrm{b}}$, whose worst-case transmission times are infinite, we can say that $p_{\varepsilon}^{\mathrm{a}}$ is more robust than $p_{\varepsilon}^{\mathrm{b}}$ if the finite-transmission-time energy set of $p_{\varepsilon}^{\mathrm{a}}$ (i.e., $\mathcal{H}_f(p_{\varepsilon}^{\mathrm{a}},E(t_0),Q(t_0))$) is larger than that of $p_{\varepsilon}^{\mathrm{b}}$ (i.e., $\mathcal{H}_f(p_{\varepsilon}^{\mathrm{b}},E(t_0),Q(t_0))$).
This is because $\mathcal{H}_f(p_{\varepsilon}^{\mathrm{a}},E(t_0),Q(t_0))$ is more likely to result in a finite transmission time in the actual transmission.
This motivates us to find the transmit power $p_{\varepsilon}$ with the largest $\mathcal{H}_f(p_{\varepsilon},E(t_0),Q(t_0))$ such that any other $\mathcal{H}_f(p_{\varepsilon}',E(t_0),Q(t_0))$ is its subset, when $\mathcal{T}^* = \infty$.

\begin{subproblem}[Energy-Set-Maximization Problem]\label{subp:Maximum-Energy-Set Problem}
Given initial battery energy $E(t_0)$, initial data queue $Q(t_0)$, and triggering threshold $\varepsilon$, if $\mathcal{T}^* = \infty$, design PTP $\tilde{p}_{n,\varepsilon}$ in each event $n$, with the knowledge of $E(t_n)$ and $Q(t_n)$, such that
\begin{align}\label{eqn:Maximum-Energy-Set Problem}
\mathcal{H}_f(p_{\varepsilon}', E(t_0), Q(t_0)) \subseteq \mathcal{H}_f(p_{\varepsilon}, E(t_0), Q(t_0)), \quad \forall p_{\varepsilon}' \in \mathfrak{P}_{\varepsilon}.
\end{align}
%
%
\end{subproblem}



As a summary of \subpref{subp:Transmission-Time-Minimization Problem} and \subpref{subp:Maximum-Energy-Set Problem}, the robust-optimal transmit power should:
\emph{achieve the minimum transmission time $\mathcal{T}^*$ under the worst-case energy-harvesting rate, if $\mathcal{T}^*$ is finite;
otherwise, ensure the largest set of $H$ that results in a finite transmission time, if $\mathcal{T}^*$ is infinite.}
Putting these two subproblems altogether, we define the Robust-Transmission-Time (RTT) problem as follows.

\begin{problem}[Robust-Transmission-Time Problem]\label{prob:Robust-Transmission-Time (RTT) Problem}
Given initial battery energy $E(t_0)$, initial data queue $Q(t_0)$, and triggering threshold $\varepsilon$, design PTP $\tilde{p}_{n,\varepsilon}$ in each event $n$, with the knowledge of $E(t_n)$ and $Q(t_n)$, such that
\begin{align}\label{eqn:RTT Problem}
\begin{cases}
p_{\varepsilon}~\mathrm{satisfies~\eqref{eqn:Transmission-Time-Minimization Problem}}, & \mathrm{if}~\mathcal{T}^* < \infty,\\
p_{\varepsilon}~\mathrm{satisfies~\eqref{eqn:Maximum-Energy-Set Problem}}, & \mathrm{if}~\mathcal{T}^* = \infty,
\end{cases}
\end{align}
where the relationship between transmit power $p_{\varepsilon}$ and the PTP $\tilde{p}_{n,\varepsilon}$ is given in~\eqref{eqn:Relationship between Transmit Power and PTP}.
\end{problem}

In the rest of this paper, we focus on how to solve the RTT problem with the event-trigger based control.

\section{Reachable Set Analysis}\label{sec:Reachable Set Analysis}

Since our proposed control method is event-trigger based (see \secref{sec:Event-Trigger Based Transmission Policy}), i.e., the transmit power is piecewise implemented in each event, in this section, we analyze all reachable battery energy and data queue (the system states) in each event $n$.
We stress that the optimal solution of the RTT problem is highly dependent on the structure of the reachable set of battery energy and data queue.

Recall that during event $n$ the TP only takes into account the information of battery energy and data queue at $t_n$ and ignores the energy-harvesting rate $H(t)$ for $t > t_n$.
The battery energy and data queue seen by the TP behave as
\begin{align}\label{eqn:Battery Energy and Data Queue From the TP Side}
\begin{cases}
\widetilde{E}_n(t) = E(t_n) - \int_{t_n}^{t} \tilde{p}_{n,\varepsilon}(\tau) \mathrm{d}\tau,\\
\widetilde{Q}_n(t) = Q(t_n) - \int_{t_n}^{t} r\left(\tilde{p}_{n,\varepsilon}(\tau)\right) \mathrm{d}\tau,
\end{cases}
\end{align}
where $\widetilde{E}_n(t)$ and $\widetilde{Q}_n(t)$ refer to the dynamics of the battery energy and data queue known by the TP based on its available information (i.e., $E(t_n)$ and $Q(t_n)$ but not $H(t)$ for $t > t_n$), which are distinct from the actual battery energy $E(t)$ and data queue $Q(t)$.\footnote{From~\eqref{eqn:Event E Happens} we know that the relationship between $\widetilde{E}_n(t)$ and $E(t)$ is $\widetilde{E}_n(t) \leq E(t) \leq \widetilde{E}_n(t) + \varepsilon$ for $t \in [t_n,~t_{n+1}]$.
The relationship between $\widetilde{Q}_n(t)$ and $Q(t)$ is $\widetilde{Q}_n(t) = Q(t)$ for $t \in [t_n,~t_{n+1}]$.}
Since in each event the PTP is designed by the TP, we should analyze the property of $\widetilde{E}_n(t)$ and $\widetilde{Q}_n(t)$ rather than $E(t)$ and $Q(t)$.
This is because the robust-optimal control of the transmit power can only be designed according to what the TP knows, i.e., $\widetilde{E}_n(t)$ and $\widetilde{Q}_n(t)$.

Now, we define the reachable set of our interest.
The reachable set contains all reachable states $\big(\widetilde{E}_n(t), \widetilde{Q}_n(t)\big)$ after implementing the PTP over the planned transmission time in the $n$\textsuperscript{th} event, i.e., all reachable $\big(\widetilde{E}_n(t_n+T_n), \widetilde{Q}_n(t_n+T_n)\big)$.
Note that the planned transmission time $T_n$ is generally different for different PTPs.

\begin{figure}
\centering
\includegraphics [width=0.6\columnwidth]{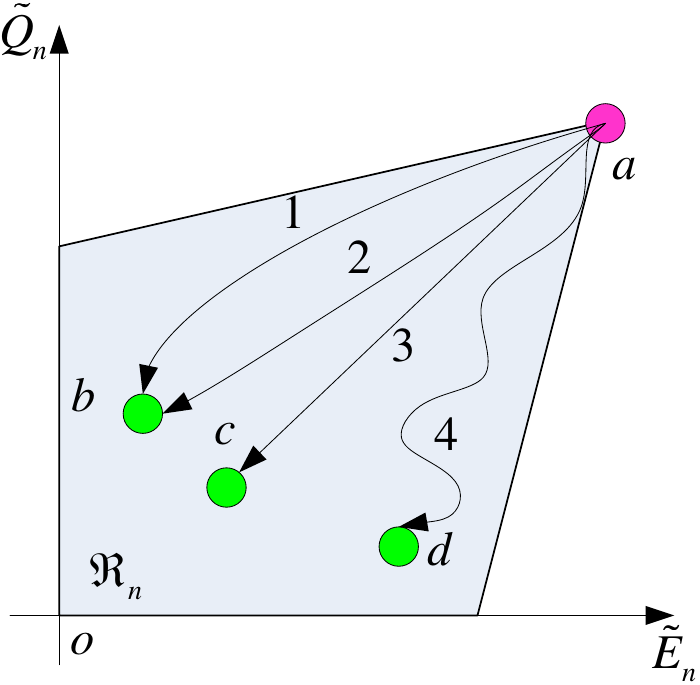}

\caption{Illustrations of reachable set in the $\widetilde{E}_n-\widetilde{Q}_n$ (two-dimensional) region for event $n$: point $a$ denotes $\left(E(t_n),Q(t_n)\right)$, and $b$, $c$ as well as $d$ are different end points $\big(\widetilde{E}_n,\widetilde{Q}_n\big)$ in the reachable set $\mathfrak{R}_n$ (blue region). The arrows with numbers $1$ and $2$ are two different paths from $a$ to $b$. The arrows with numbers $3$ and $4$ are the paths from $a$ to $c$ and $d$, respectively.}\label{fig:Sketch Map for Reachable Set}
\end{figure}

\figref{fig:Sketch Map for Reachable Set} gives a pictorial illustration of the reachable set:
At $t = t_n$, the system state is at point $\big(\widetilde{E}_n(t_n), \widetilde{Q}_n(t_n)\big)$ (i.e., point $a$).
After $t_n$, the PTP pushes the system state (seen by the TP) to move along the arrow (different PTPs correspond to different arrows).
At $t = t_n + T_n$, the state stops at a point (e.g., point $b$) which corresponds to $\big(\widetilde{E}_n(t_n+T_n), \widetilde{Q}_n(t_n+T_n)\big)$.
All $\big(\widetilde{E}_n(t_n+T_n), \widetilde{Q}_n(t_n+T_n)\big)$ compose the reachable set (i.e., the shaded area).
%
%
We define the reachable set as follows.

\begin{definition}[Reachable Set]\label{def:Reachable Set}
From given $E(t_n)$ and $Q(t_n)$, the reachable set in $n$\textsuperscript{th} event is
\begin{align*}
\begin{split}
\mathfrak{R}_n = &\Big\{\left(\widetilde{E}_n,\widetilde{Q}_n\right):\widetilde{E}_n = E(t_n) - \int_{t_n}^{t_n+T_n}\tilde{p}_{n,\varepsilon}(\tau)d\tau \geq 0,\\
&\widetilde{Q}_n = Q(t_n) - \int_{t_n}^{t_n+T_n}r(\tilde{p}_{n,\varepsilon}(\tau))d\tau \geq 0,\\
&\tilde{p}_{n,\varepsilon} \in \widetilde{\mathfrak{P}}_{n,\varepsilon},~T_n < \infty\Big\},
\end{split}
\end{align*}
where we use $\big(\widetilde{E}_n,\widetilde{Q}_n\big)$ rather than $\big(\widetilde{E}_n(t),\widetilde{Q}_n(t)\big)$ to represent the point in $\mathfrak{R}_n$, as $\big(\widetilde{E}_n(t_n+T_n),\widetilde{Q}_n(t_n+T_n)\big)$ can be the same with different $T_n$, which violates the definition of set.
\end{definition}

In \figref{fig:Sketch Map for Reachable Set}, we can see that different PTPs correspond to different paths or arrows in the figure, which may or may not arrive at the same end point.

Even though Definition~\ref{def:Reachable Set} gives an expression of the reachable set, it is too abstract and not convenient for design.
To give a more explicit form of reachable set, we define the Rate-Power Equilibrium (RPE) to help the subsequent analysis.
%

\begin{definition}[Rate-Power Equilibrium (RPE)]\label{def:Rate-Power Equilibrium}
The rate-power line is defined in the $r-p$ plane (see~\figref{fig:Rate-power line, rate function, RPE and IRPE}):
\begin{equation}\label{eqn:Rate-Power Line}
r = K_n p,~\textmd{where}~K_n = \frac {Q(t_n) - \widetilde{Q}_n} {E(t_n) - \widetilde{E}_n}.
\end{equation}
The intersection of rate-power line and rate function $r = \log_2(1 + p)$ for $p \in (0,~p_{max}]$ (here, $p$ is a scalar) is called the RPE, and the corresponding transmit power of the RPE is labeled by $p_n^e$.
For $\widetilde{E}_n = E(t_n)$ or $\widetilde{Q}_n = Q(t_n)$, we define their $p_n^e = 0$, even though no RPE exists.
%
\end{definition}


From \defref{def:Rate-Power Equilibrium}, we see that any arbitrary pair of values $\big(\widetilde{E}_n, \widetilde{Q}_n\big)$ has a corresponding rate-power line in the $r-p$ plane.
Because of the concavity of the rate function, there exists at most one RPE for an arbitrary $\big(\widetilde{E}_n, \widetilde{Q}_n\big)$.
If the RPE exists, we can use the following remark to calculate it.

\begin{remark}\label{rek:Closed-Form RPE}
Solving $K_n p_n^e = \log_2(1 + p_n^e)$, we have
\begin{align}\label{eqn:P_RPE for Gaussian Channel}
p_n^e = - \frac {1} {K_n \ln 2} W_{-1}\left(- K_n \ln 2 \cdot 2^{-K_n} \right)  - 1,
\end{align}
where $W_{-1}$ is the real valued Lambert W function~\cite{CorlessR1996} in the lower branch ($W \leq -1$).
Here the RPE is the point $(p_n^e,\log_2(1 + p_n^e))$ in the $r-p$ plane.
\end{remark}

The following lemma makes the link between the reachable set and the RPE, which helps us to find an explicit expression of reachable set in order to facilitate transmit power design.

\begin{figure}
\centering
\includegraphics [width=0.7\columnwidth]{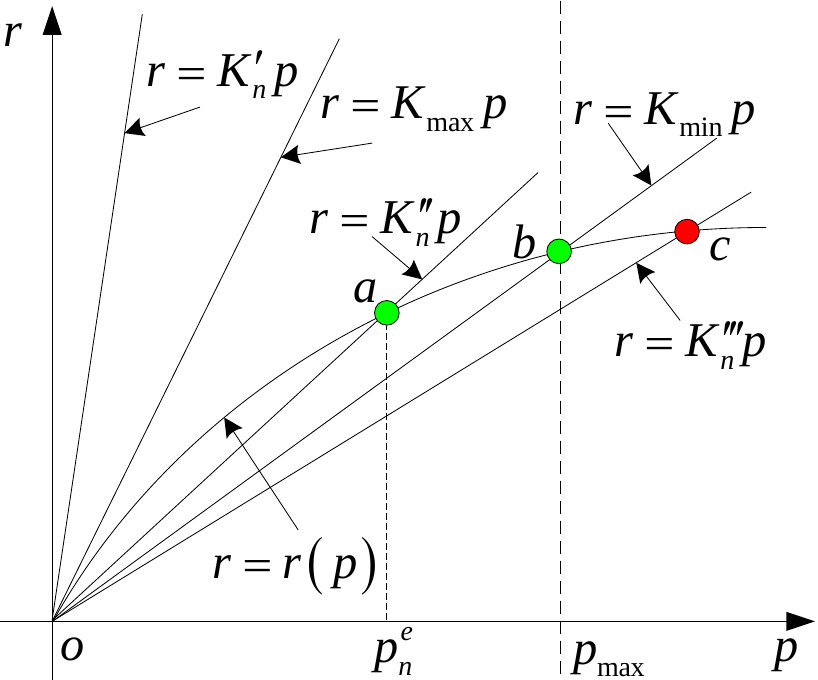}

\caption{Rate-power line, rate function and RPE, where $r(p) = \log_2(1 + p)$ and $K'_n > K_{\max} > K''_n > K_{\min} > K'''_n$.
For $K'_n$ and $K_{\max}$, no positive intersections exist due to the large $K_n$.
For $K''_n$, $K_{\min}$ and $K'''_n$, positive intersections exist because of the small $K_n$:
Points $a$ as well as $b$ are RPEs, but point $c$ is not a RPE due to $p_n^e > p_{\max}$.
We can see that $p_n^e$ decreases as $K_n$ going large.}\label{fig:Rate-power line, rate function, RPE and IRPE}
\end{figure}

\begin{lemma}[Criterion on Points in Reachable Set]\label{lem:Criterion on Points in Reachable Set}
\begin{align}\label{eqn:Points in Reachable Set Other Than starting point}
\left(\widetilde{E}_n,\widetilde{Q}_n\right) \in \mathfrak{R}_n\backslash\{(E(t_n),Q(t_n))\}
\end{align}
if and only if the RPE for $\big(\widetilde{E}_n,\widetilde{Q}_n\big)$ exists.
\end{lemma}

\begin{IEEEproof}
See \apxref{apx:Proof of Lemma 1}.
\end{IEEEproof}

\lemref{lem:Criterion on Points in Reachable Set} tells that except for $(E(t_n),Q(t_n))$, any point in reachable set has a RPE, and any point which has a RPE must be in the reachable set.
%
%
Based on \lemref{lem:Criterion on Points in Reachable Set}, an explicit expression of reachable set can be given.

\begin{proposition}[Expression for Reachable Set]\label{prop:Expressions of Reachable Set}
Given $(E(t_n),Q(t_n))$, the reachable set satisfies
\begin{multline}\label{eqn:Expressions of Reachable Set}
\mathfrak{R}_n\backslash\{\left(E(t_n),Q(t_n)\right)\} \!= \!\left\{\left(\widetilde{E}_n,\widetilde{Q}_n\right): \! K_{\min} \! \leq \! K_n \! < \! K_{\max},\right.\\
\left.0 \! \leq \! \widetilde{E}_n \!<\! E(t_n), 0 \leq \! \widetilde{Q}_n\! \!< Q(t_n)\right\},
\end{multline}
where $K_n$ is a function of $\widetilde{E}_n$ and $\widetilde{Q}_n$ given in~\eqref{eqn:Rate-Power Line}, and
\begin{align}\label{eqn:Kmin and Kmax}
K_{\min} := \frac {r(p_{\max})} {p_{\max}},~K_{\max} := \lim_{x\rightarrow0^+} \frac {r(x)} {x} = \frac {1} {\ln 2}.
\end{align}
\end{proposition}

\begin{IEEEproof}
See \apxref{apx:Proof of Proposition 1}.
\end{IEEEproof}

\propref{prop:Expressions of Reachable Set} means that the point $\big(\widetilde{E}_n,\widetilde{Q}_n\big)$ is in the reachable set if and only if the corresponding $K_n$ (i.e., the slope of the corresponding rate-power line) is within a certain range.
As shown in \figref{fig:Rate-power line, rate function, RPE and IRPE}, the slope $K_n$ decreases as $p_n^e$ grows, which implies:
On the one hand, $K_{\max}$ is the supremum of the slope $K_n$ to have an intersection between the rate-power line and the rate function (i.e., to have a RPE).
On the other hand, due to the maximum power constraint, $K_{\min}$ is the minimum slope to have a RPE.

\begin{figure*}[htb]
\centering
\subfigure[]{\includegraphics [width=0.6\columnwidth, trim = 0 0 0 0]{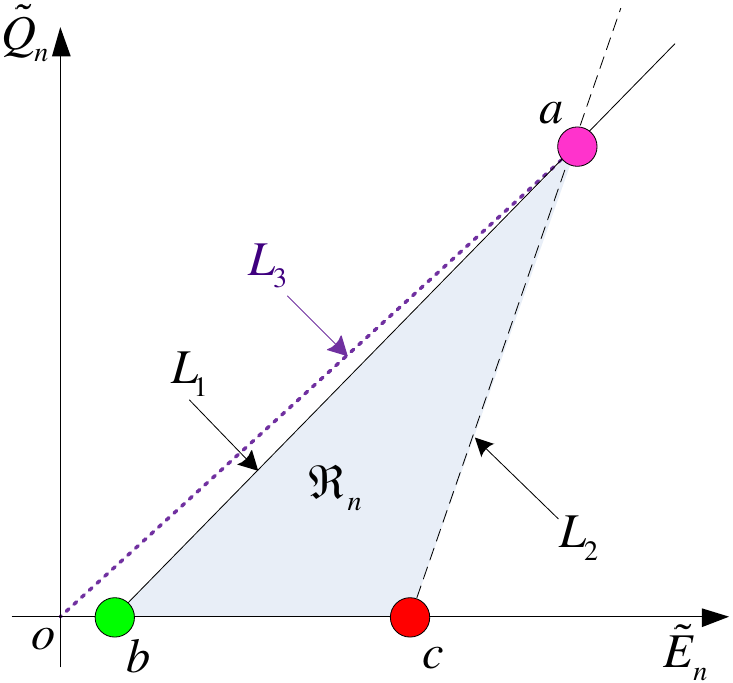}\label{fig:Reachable Set Battery Energy Surplus}}\hspace{0.1em}
\subfigure[]{\includegraphics [width=0.6\columnwidth, trim = 0 0 0 0]{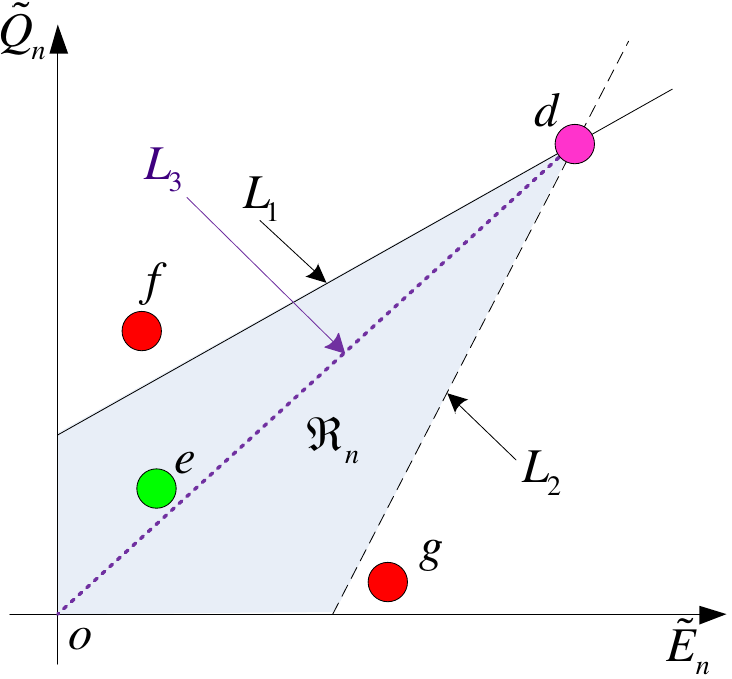}\label{fig:Reachable Set Battery Energy Data Queue Balance}}\hspace{0.1em}
\subfigure[]{\includegraphics [width=0.6\columnwidth, trim = 0 0 0 0]{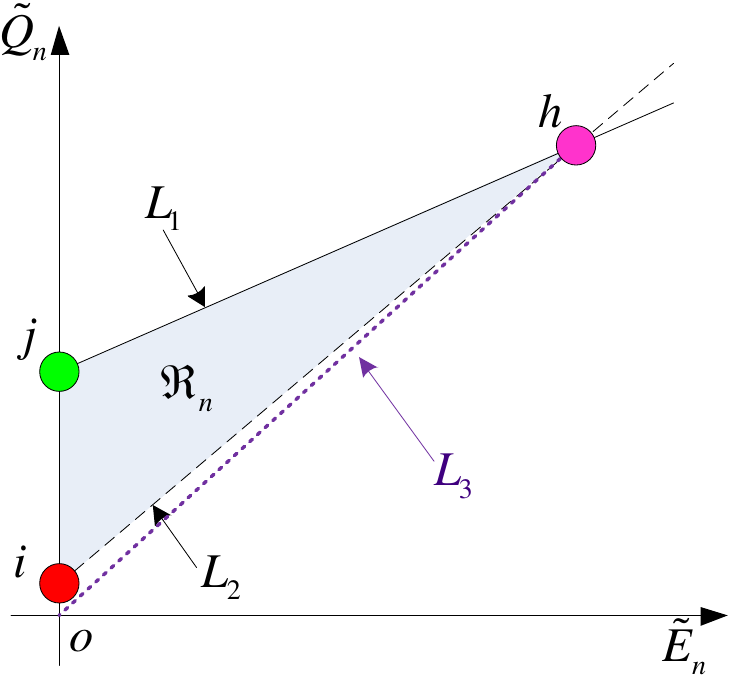}\label{fig:Reachable Set Data Queue Surplus}}

\caption{Shapes of 3 cases of reachable sets (including the starting points $a,d,h$), where $L_1$, $L_2$, and $L_3$ correspond to lines $\widetilde{Q}_n - Q(t_n) = K_{\min} \big(\widetilde{E}_n - E(t_n)\big)$, $\widetilde{Q}_n - Q(t_n) = K_{\max} \big(\widetilde{E}_n - E(t_n)\big)$, and $\widetilde{Q}_n - Q(t_n) = K_{\mathrm{bal}} \big(\widetilde{E}_n - E(t_n)\big)$, respectively.
(a) Energy-abundant case ($K_n^{\mathrm{bal}} < K_{\min}$), point $b$ is in the reachable set, whereas point $c$  is not in the reachable set.
(b) Energy-balanced case ($K_{\min} \leq K_n^{\mathrm{bal}} < K_{\max}$), point $e$ is in the reachable set, while points $f$ and $g$ are not in the reachable set.
(c) Energy-scarce case ($K_n^{\mathrm{bal}} \geq K_{\max}$), point $j$  is in the reachable set, but point $i$ is not in the reachable set.}\label{fig:Reachable Sets}
\end{figure*}

%
A pictorial illustration of the reachable set is shown in \figref{fig:Reachable Sets}, and it can be easily categorized into three cases, depending on the relationship among $K_{\min}$, $K_{\max}$, and $K_n^{\mathrm{bal}}$ (called the energy-balanced slope), where
\begin{align}\label{eqn:Kbalance}
K_n^{\mathrm{bal}} := \frac {Q(t_n) - 0} {E(t_n) - 0} = \frac {Q(t_n)} {E(t_n)}.
\end{align}
To be more specific, \figref{fig:Reachable Set Battery Energy Surplus}, \figref{fig:Reachable Set Battery Energy Data Queue Balance}, and \figref{fig:Reachable Set Data Queue Surplus} correspond to $K_n^{\mathrm{bal}} < K_{\min}$, $K_{\min} \leq K_n^{\mathrm{bal}} < K_{\max}$ and $K_n^{\mathrm{bal}} \geq K_{\max}$, respectively.
These three cases have important physical meanings, given in the following remark.

\begin{remark}[Categorization of Reachable Sets]\label{rek:Illustration on Fig. Reachable Sets}
In Fig.~\ref{fig:Reachable Set Battery Energy Surplus}, $E_n$ is always greater than $0$, $\forall (\widetilde{E}_n,\widetilde{Q}_n) \in \mathfrak{R}_n$, which implies that the battery energy is abundant. This means that when the data queue is cleared, there is still battery energy remaining, no matter what PTP is used.
In this case, we say that $\mathfrak{R}_n$ is energy-abundant.
In Fig.~\ref{fig:Reachable Set Battery Energy Data Queue Balance}, origin $o$ is in the reachable set, which means the data queue can be cleared by using all the energy stored in the battery, and in this case, we say that $\mathfrak{R}_n$ is energy-balanced.
In Fig.~\ref{fig:Reachable Set Data Queue Surplus}, $Q_n$ is always greater than $0$, $\forall (\widetilde{E}_n,\widetilde{Q}_n) \in \mathfrak{R}_n$, which means the data queue cannot be cleared with the available battery energy, no matter what PTP is employed.
In this case, we say that $\mathfrak{R}_n$ is energy-scarce.
\end{remark}

The RPE not only helps to shape the reachable set (see \propref{prop:Expressions of Reachable Set}), but also gives the time-optimal PTP.
Recall that from one starting point $(E(t_n),Q(t_n))$, there are multiple PTPs that reach the same end point $\left(\widetilde{E}_n,\widetilde{Q}_n\right)$ (see \figref{fig:Sketch Map for Reachable Set}).
These PTPs, however, spend different amount of planned transmission time $T_n$.
Hence, we need to find the time-optimal PTP that has the minimum $T_n$ for each point $\left(\widetilde{E}_n,\widetilde{Q}_n\right) \in \mathfrak{R}_n$.

For a given starting point $(E(t_n),Q(t_n))$ and an end point $\big(\widetilde{E}_n,\widetilde{Q}_n\big)$, the planned transmission time $T_n:\widetilde{\mathfrak{P}}_{n,\varepsilon} \rightarrow \overline{\mathbb{R}}_+$ is a non-negative functional of the PTP, and the time-optimal PTP has the smallest $T_n$ (and we mark the minimum planned transmission time as $\underline{T}_n\big[\widetilde{E}_n,\widetilde{Q}_n\big]$ for end point $\big(\widetilde{E}_n,\widetilde{Q}_n\big)$), i.e.,
\begin{align}\label{eqn:Optimal Time}
\underline{T}_n\big[\widetilde{E}_n,\widetilde{Q}_n\big] = \underset {\tilde{p}_{n,\varepsilon} \in \widetilde{\mathfrak{P}}_{n,\varepsilon}\big[\widetilde{E}_n,\widetilde{Q}_n\big]} {\inf} T_n\left(\tilde{p}_{n,\varepsilon}\right),
\end{align}
where $\widetilde{\mathfrak{P}}_{n,\varepsilon}\big[\widetilde{E}_n,\widetilde{Q}_n\big]$ stands for those PTPs to make the end point as $\big(\widetilde{E}_n,\widetilde{Q}_n\big)$.
The following proposition shows that the transmit power in the time-optimal PTP for a given pair of starting point and end point is unique and remains constant at the value of $p_n^e$ (the transmit power corresponding to the RPE) over the planned transmission time $T_n$.

\begin{proposition}[Time-Optimal PTP]\label{prop:Time-Optimal PTP}
$\forall \big(\widetilde{E}_n,\widetilde{Q}_n\big) \in \mathfrak{R}_n\backslash\{(E(t_n),Q(t_n))\}$, the unique time-optimal PTP to achieve $\underline{T}_n\big[\widetilde{E}_n,\widetilde{Q}_n\big]$ is $\tilde{p}_{n,\varepsilon}^{\mathrm{TIO}}$ with parameters
\begin{align}\label{eqn:Time-Optimal PTP}
(\tilde{p}_{n,\varepsilon}(t), T_n) = \left(p_n^e, \underline{T}_n\big[\widetilde{E}_n,\widetilde{Q}_n\big]\right),
\end{align}
where $p_n^e$ is the transmit power of the corresponding RPE which can be calculated by~\eqref{eqn:P_RPE for Gaussian Channel} in \rekref{rek:Closed-Form RPE}, and the minimum planned transmission time is:
\begin{align}\label{eqn:Optimal Planned Transimission Time}
\underline{T}_n \big[\widetilde{E}_n,\widetilde{Q}_n\big] = \frac {Q(t_n) - \widetilde{Q}_n}{r\left(p_n^e\right)} = \frac {E(t_n) - \widetilde{E}_n}{p_n^e}.
\end{align}
\end{proposition}

\begin{IEEEproof}
See \apxref{apx:Proof of Proposition 2}.
\end{IEEEproof}

With \propref{prop:Time-Optimal PTP}, we can calculate the time-optimal PTP through~\eqref{eqn:Time-Optimal PTP} to minimize the planned transmission time $\underline{T}_n\big[\widetilde{E}_n,\widetilde{Q}_n\big]$ for any point in reachable set except for $(E(t_n),Q(t_n))$.
But obviously, the optimal time for $(E(t_n),Q(t_n))$ is $\underline{T}_n[E(t_n),Q(t_n)] = 0$.
Note that different end points $\big(\widetilde{E}_n,\widetilde{Q}_n\big)$ correspond to different $\underline{T}_n[E(t_n),Q(t_n)]$.
If the TP wants to clear the data queue with a minimum planned transmission time, it is equivalent to consider the end points with $\widetilde{Q}_n = 0$ and select one from them which has the minimum $\underline{T}_n\big[\widetilde{E}_n,\widetilde{Q}_n\big]$.
This result is given in \corref{cor:Minimum Transmission-Time to Clear the Data Queue from the TP Side}.

\begin{corollary}\label{cor:Minimum Transmission-Time to Clear the Data Queue from the TP Side}
If $K_n^{\mathrm{bal}} < K_{\min}$, we have
\begin{align}\label{eqnincor: Energy-Abundant Case}
\!\!\!\!\!\!\argmin_{\big(\widetilde{E}_n,\widetilde{Q}_n\big) \in \mathfrak{R}_n, \widetilde{Q}_n = 0}\!\!\underline{T}_n \big[\widetilde{E}_n,\widetilde{Q}_n\big] = \left(E(t_n) - \frac{Q(t_n)}{K_{\min}}, 0\right),
\end{align}
whose $p_n^e$ (see \propref{prop:Time-Optimal PTP}) is $p_{\max}$.
If $K_{\min} \leq K_n^{\mathrm{bal}} < K_{\max}$, we have
\begin{align}\label{eqnincor: Energy-Balanced Case}
\argmin_{\big(\widetilde{E}_n,\widetilde{Q}_n\big) \in \mathfrak{R}_n, \widetilde{Q}_n = 0}\underline{T}_n \big[\widetilde{E}_n,\widetilde{Q}_n\big] = (0, 0),
\end{align}
whose $p_n^e$ is labeled as $p_n^{\mathrm{bal}}$, which has the following form:
\begin{align}\label{eqn:pbaln}
p_n^{\mathrm{bal}} = - \frac {1} {K_n^{\mathrm{bal}} \ln 2} W_{-1}\left(- K_n^{\mathrm{bal}} \ln 2 \cdot 2^{-K_n^{\mathrm{bal}}} \right)  - 1,
\end{align}
where $K_n^{\mathrm{bal}}$ is given in~\eqref{eqn:Kbalance} and $W_{-1}$ is the real valued Lambert W function in the lower branch~\cite{CorlessR1996}.
If $K_n^{\mathrm{bal}} \geq K_{\max}$, we cannot find any end point with $\widetilde{Q}_n = 0$.
\end{corollary}

\begin{IEEEproof}
See \apxref{apx:Proof of Corollary 1}.
\end{IEEEproof}

\begin{remark}\label{rek:Minimal-Time to Clear the Data Queue from the TP Side}
We claim that \corref{cor:Minimum Transmission-Time to Clear the Data Queue from the TP Side} plays an important role in the solution to the RTT problem, which is shown in \thmref{thm:Optimal Solution to RTT Problem}.
More details can be found in the proof of \thmref{thm:Optimal Solution to RTT Problem} (see \apxref{apx:Proof of Theorem 1}).
Briefly speaking, the time-optimal PTPs corresponding to $p_n^e$ in \corref{cor:Minimum Transmission-Time to Clear the Data Queue from the TP Side} give the solution to the RTT problem for cases $K_n^{\mathrm{bal}} < K_{\min}$ and $K_{\min} \leq K_n^{\mathrm{bal}} < K_{\max}$.
\end{remark}

\section{Solution to RTT Problem and Discussion on Triggering Condition}\label{sec:Solution to RTT Problem and Discussion on Triggering Condition}

The analysis on the reachable set of battery energy and data queue in an arbitrary event as well as the result on time-optimal PTP have enabled us to solve the RTT problem defined in \probref{prob:Robust-Transmission-Time (RTT) Problem}.
In this section, firstly, we present the optimal solution of the RTT problem, and then discuss the effect of the triggering threshold $\varepsilon$.

\subsection{Optimal Solution to RTT Problem}\label{sec:Optimal Solution to RTT Problem}

\begin{theorem}[Optimal Solution to RTT Problem]\label{thm:Optimal Solution to RTT Problem}
The optimal solution of RTT problem is $\tilde{p}_{n,\varepsilon}^\mathrm{R}$ with the parameters\footnote{Recall that any $\tilde{p}_{n,\varepsilon}$ in~\eqref{eqn:Planned Transmission Policy} can be determined by two parameters $\tilde{p}_{n,\varepsilon}(t)$ and $T_n$.}:
\begin{align}\label{eqn:Optimal Solution of RTT Problem}
\left(\tilde{p}_{n,\varepsilon}(t),T_n\right) =
\begin{cases}
\left(p_{\max}, \frac{Q(t_n)}{r(p_{\max})}\right) & K_n^{\mathrm{bal}} < K_{\min},\\
\left(p_n^{\mathrm{bal}}, \frac{E(t_n)}{p_n^{\mathrm{bal}}}\right) & K_{\min} \leq K_n^{\mathrm{bal}} < K_{\max},\\
\left(0, 0\right) & K_n^{\mathrm{bal}} \geq K_{\max},
\end{cases}
\end{align}
where $p_n^{\mathrm{bal}}$ is given in~\eqref{eqn:pbaln}.
The corresponding actual transmit power implemented by $\tilde{p}_{n,\varepsilon}^\mathrm{R}$ is labeled as $p_{\varepsilon}^\mathrm{R}$.
\end{theorem}

\begin{IEEEproof}
See \apxref{apx:Proof of Theorem 1}.
%
\end{IEEEproof}

Recall the event-trigger based framework for transmission design summarized in \algref{alg:Event-Trigger Based Control}.
The solution to the design problem in Lines~\ref{line:Design PTP 1} and~\ref{line:Design PTP 2} is now given in \thmref{thm:Optimal Solution to RTT Problem}.

\begin{remark}
The structure of $\tilde{p}_{n,\varepsilon}^\mathrm{R}$ is easy to understand.
The first row corresponds to the energy-abundant case for $\mathfrak{R}_n$ (see \figref{fig:Reachable Set Battery Energy Surplus}), and in this case, the maximum power is used to transmit, since there is enough energy.
Likewise, the second row stands for the energy-balanced case (see \figref{fig:Reachable Set Battery Energy Data Queue Balance}), and the corresponding transmit power is $p_n^{\mathrm{bal}}$, which can clear the data queue and use up the battery energy at the same time.
In the third row, the energy-scarce case (see \figref{fig:Reachable Set Data Queue Surplus}), the transmitter sends nothing, which can be explained as that any transmission in this case would make things worse.
\end{remark}

\begin{remark}
The worst-case energy-harvesting rate for the optimal transmit power design $p_{\varepsilon}^\mathrm{R}$ exists and is given by $H_{p_{\varepsilon}^\mathrm{R}}: t \mapsto 0$, i.e., no energy arrival in $[t_0,~\infty)$ (the proof is given in \lemref{lem:The Worst-Case Energy-Harvesting Rate of Designed PTP} in \apxref{apx:Proof of Theorem 1}).
We denote such no-energy-arrival case as $H_o$.
It should be noted that the worst-case energy-harvesting rate of any given transmission power function is not always $H_o$.
Indeed, to determine the worst-case energy-harvesting rate of a given transmission power design is difficult in general, which is the main difficulty in solving the RTT problem.
In \secref{sec:Simulation Results}, we will show an example of a transmit power design of which the worst-case energy-harvesting rate is very different from $H_o$.
\end{remark}

\subsection{Discussions on the Triggering Threshold}\label{sec:Discussions on Triggering Condition}

In \secref{sec:Optimal Solution to RTT Problem}, the optimal solution of the RTT problem is investigated for a given triggering threshold.
A natural question is that: how does the triggering threshold $\varepsilon$ affect the system behavior (i.e., $\mathcal{T}^*$ and $\mathcal{H}_f(p_{\varepsilon}^\mathrm{R}, E(t_0), Q(t_0))$)?
In this subsection, we give the corresponding answers.

Firstly, we show that when $\mathcal{T}^* < \infty$ holds, $\mathcal{T}^*$ is independent of the value of $\varepsilon$, which is easy to verify, since the worst-case $H$ of the robust-optimal solution $\tilde{p}_{n,\varepsilon}^\mathrm{R}$ is $H_o$.
In this worst case, $\forall \varepsilon_1, \varepsilon_2 > 0$, we have $p_{\varepsilon_1}^\mathrm{R} = p_{\varepsilon_2}^\mathrm{R}$.
This is because: For $H_o$, the next event would never be triggered.
As a result, the actual transmit power is only dependent on the PTP designed in event $0$ which is not affected by the value of the triggering threshold.

Different from the $\mathcal{T}^* < \infty$ case, for the $\mathcal{T}^* = \infty$ case, $\mathcal{H}_f(p_{\varepsilon}^\mathrm{R}, E(t_0), Q(t_0))$ is dependent on the value of $\varepsilon$, and the following proposition tells that the smaller $\varepsilon$ is, the larger $\mathcal{H}_f(p_{\varepsilon}^\mathrm{R}, E(t_0), Q(t_0))$ will be.
%

\begin{proposition}\label{prop:The Inclusion Property of Higher Resolution}
For $\mathcal{T}^* = \infty$, if $\varepsilon^b$ is a multiple of $\varepsilon^a$ with multiplier $z \in \mathbb{Z}_+ \setminus \{1\}$, i.e., $z \varepsilon^a = \varepsilon^b$, then $\mathcal{H}_f(p_{\varepsilon^a}^{\mathrm{R}}, E(t_0), Q(t_0)) \supset \mathcal{H}_f(p_{\varepsilon^b}^{\mathrm{R}}, E(t_0), Q(t_0))$.
\end{proposition}

\begin{IEEEproof}
See \apxref{apx:Proof of Proposition 3}.
\end{IEEEproof}

\propref{prop:The Inclusion Property of Higher Resolution} implies that the smaller $\varepsilon$ is, the larger the set $\mathcal{H}_f(p_{\varepsilon}^{\mathrm{R}}, E(t_0), Q(t_0))$ will be, which means the more cases of energy-harvesting rate $H$ can result in a finite transmission time.
However, due to the limited computational resource, we cannot make $\varepsilon$ arbitrarily small because smaller $\varepsilon$ leads to more frequent event triggers.
In practice, we should balance the computational accuracy and efficiency.

\section{Simulation Results}\label{sec:Simulation Results}

Now, we present the simulation results to illustrate the benefit of the proposed transmission designs based on robust-optimal control.

Since there are no comparable online algorithms in the literature, we take the following three designs in the event-trigger control framework as examples to compare with our design.
The first is an estimation-based algorithm, labeled as $p_{\varepsilon}^{\mathrm{E}}$, which estimates the future energy-harvesting rate based on the energy-harvesting rate in the past.
The corresponding PTP $\tilde{p}_{n,\varepsilon}^\mathrm{E}$ in each event $n$ has parameters:
\begin{multline}\label{eqn:Estimation-Based PTP}
\left(\tilde{p}_{n,\varepsilon}(t),T_n\right) =\\
\begin{cases}
\left(p_{\max}, \frac{Q(t_n)}{r(p_{\max})}\right) & K_n^{\mathrm{bal}} < K_{\min},\\
\left(p_n^{\mathrm{bal}}+\vartriangle\!\!p, \frac{E(t_n)}{p_n^{\mathrm{bal}}+\vartriangle p}\right) & K_{\min} \leq K_n^{\mathrm{bal}} < K_{\max},\\
\left(0, 0\right) & K_n^{\mathrm{bal}} \geq K_{\max},
\end{cases}
\end{multline}
where $\vartriangle\!\!p = 0$ for $n = 0$, and $\vartriangle\!\!p = [E(t_n) - E(t_{n-1})]/(t_n - t_{n-1})$ for $n > 0$.
Compared with the robust-optimal PTP design in~\eqref{eqn:Optimal Solution of RTT Problem}, $\tilde{p}_{n,\varepsilon}^\mathrm{E}$ has the same structure and the only difference is the second line in~\eqref{eqn:Estimation-Based PTP}, where $\vartriangle\!\!p$ is the additionally allocated power in every event.
We can see that $\vartriangle\!\!p$ is dependent on the average energy-harvesting rate in the last event, i.e., $[E(t_n) - E(t_{n-1})]/(t_n - t_{n-1})$.
The additionally allocated power takes into account the future energy arrival whose rate is estimated to be the same as that in the last event.
In contrast, the robust-optimal design does not assume any future energy arrival within the current event.
Therefore, the estimation-based design is smarter than our robust-optimal design $p_{\varepsilon}^{\mathrm{R}}$ when $H$ happens to be a stationary process, since the average energy-harvesting rate $[E(t_n) - E(t_{n-1})]/(t_n - t_{n-1})$ in the past can be a reasonably accurate estimate of the energy-harvesting rate in the future.
The second design is the modified estimation-based algorithm, labeled as $p_{\varepsilon}^{\mathrm{M}}$, which has the same structure~\eqref{eqn:Estimation-Based PTP} to $p_{\varepsilon}^{\mathrm{E}}$, but with a different $\vartriangle\!\!p$ that: $\vartriangle\!\!p = 0$ for $n = 0$, and $\vartriangle\!\!p = 0.25[E(t_n) - E(t_{n-1})]/(t_n - t_{n-1})$ for $n > 0$.
The modified estimation-based algorithm $p_{\varepsilon}^{\mathrm{M}}$ is more conservative than $p_{\varepsilon}^{\mathrm{E}}$ because only a quarter of the estimated energy is used.
The third design is the greedy algorithm, labeled by $p_{\varepsilon}^{\mathrm{G}}$, which simply transmits data with maximum power in every event $n$ if $E(t_n) > 0$ and $Q(t_n) > 0$.

We present simulation results in three scenarios.
In the first two scenarios, the energy-harvesting rates are deterministic but totally unknown to the transmitter, and in the third scenario, the energy-harvesting rate is a non-stationary stochastic process whose statistics are totally unknown to the transmitter.

In the section, the units of all parameters are normalized, and we assume $t_0 = 0$.

\begin{figure}[h]
\centering
\subfigure[]{\includegraphics [width=0.7\columnwidth]{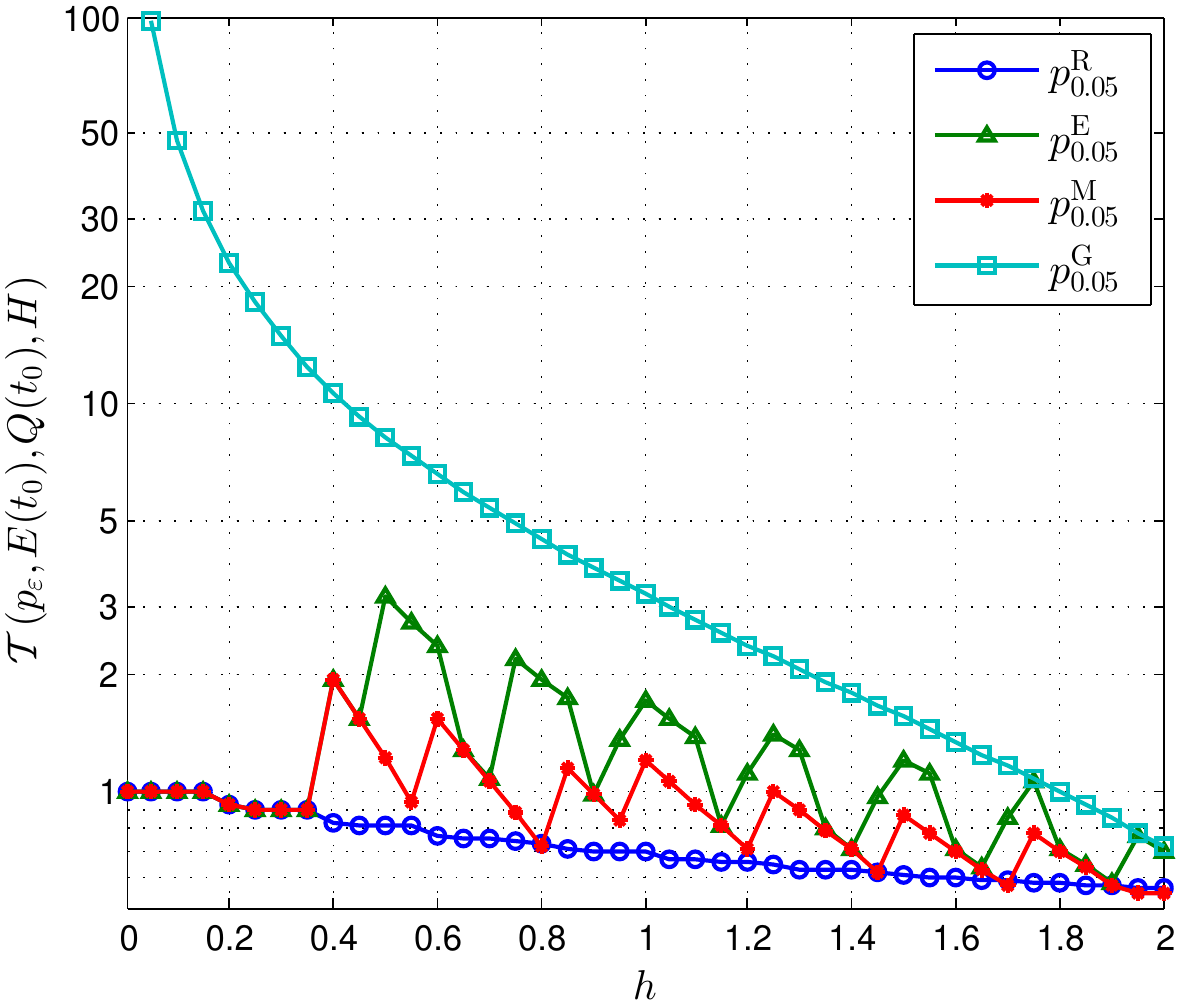}\label{fig:TTO Problem Simulations T Star is Finite}}\\
\subfigure[]{\includegraphics [width=0.7\columnwidth]{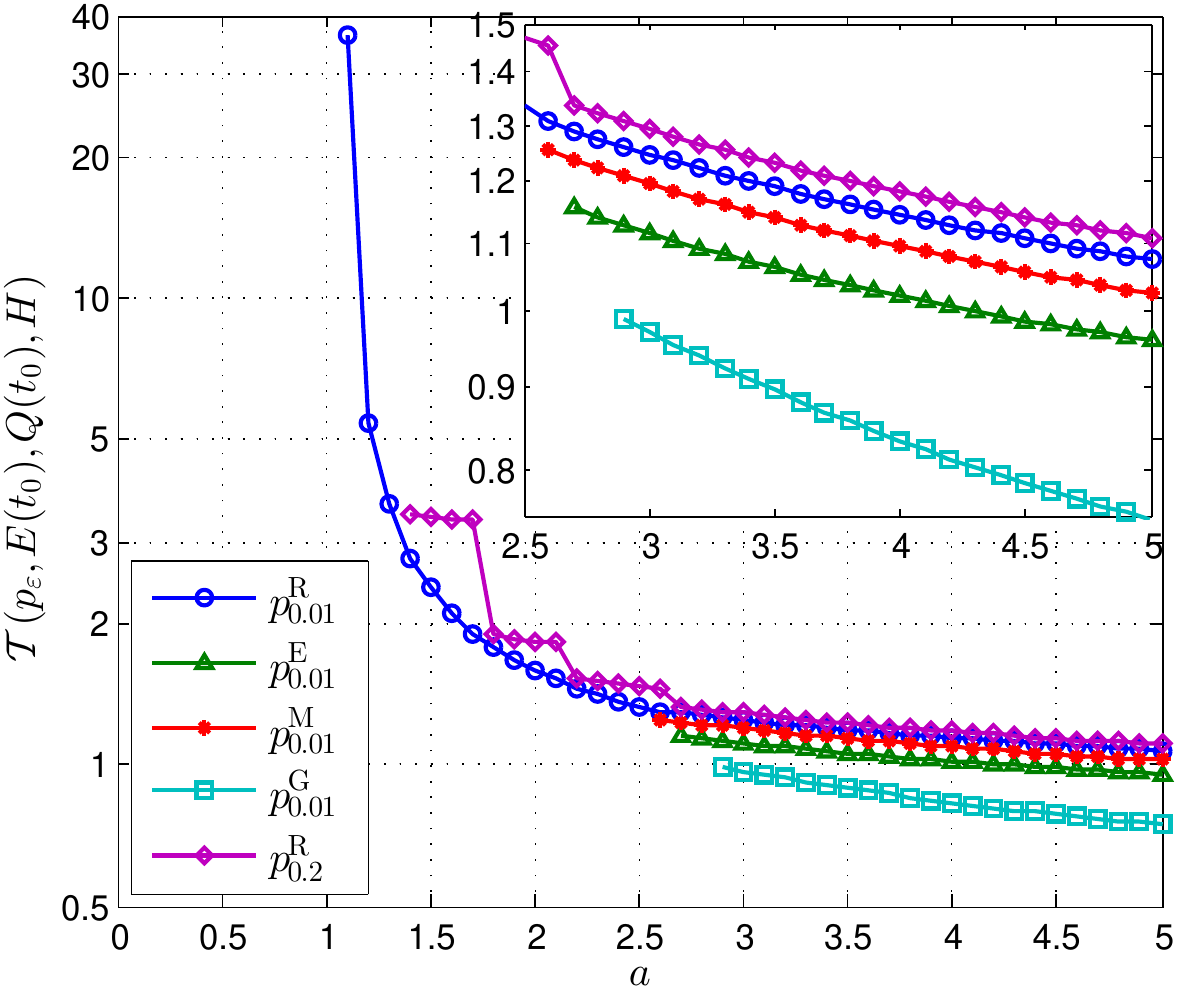}\label{fig:TTO Problem Simulations T Star is Infinite}}

\caption{Transmission time comparisons:
(a) $\mathcal{T}^* < \infty$;
(b) $\mathcal{T}^* = \infty$.
}\label{fig:TTO Problem Simulations}
\end{figure}

\textbf{Scenario 1.}
We set the initial battery energy as $E(0) = 1$, the initial data queue length as $Q(0) = 1$, the maximum transmit power as $p_{\max} = 3$, and the triggering threshold as $\varepsilon = 0.05$.
The energy-harvesting rate is chosen as $H(t) = h$ for $t \in [0,~0.2)$, and $H(t) = h/10$ for $t \in [0.2,~\infty)$, where $h$ is selected from $[0,~2]$.
The transmission time comparison of $p_{0.05}^{\mathrm{R}}$, $p_{0.05}^{\mathrm{E}}$, $p_{0.05}^{\mathrm{M}}$, and $p_{0.05}^{\mathrm{G}}$ is given in \figref{fig:TTO Problem Simulations T Star is Finite}, from which we can observe that:
\begin{itemize}
\item[1)]   The worst transmission time of $p_{0.05}^{\mathrm{R}}$ is $1$.
            The worst-case energy harvesting rate is $H(t)$ with parameter $h \in [0,~0.15]$, hence including $H_o$ as a worst case.
\item[2)]   The worst transmission time of $p_{0.05}^{\mathrm{E}}$ in this figure is $3.2$.
            The worst-case energy harvesting rate is with parameter $h = 0.5$.
            Clearly, $H_o$ is not the worst case.
\item[3)]   The worst transmission time of $p_{0.05}^{\mathrm{M}}$ in this figure is $1.95$.
            The worst-case energy harvesting rate is with parameter $h = 0.5$.
            Clearly, $H_o$ is not the worst case.
\item[4)]   The worst transmission time of $p_{0.05}^{\mathrm{G}}$ is $\infty$.
            The worst-case energy harvesting rate is $H_o$.
%
\end{itemize}
We can see that the transmission time is guaranteed by our design to be not greater than $1$, while the other three designs can result in a transmission time much larger than $1$.

\textbf{Scenario 2.}
We set the initial battery energy as $E(0) = 0.2$, the initial data queue length as $Q(0) = 1$, and the maximum transmit power as $p_{\max} = 3$.
The energy-harvesting rate is chosen as $H(t) = a |\sin t|$, for $t \in [0,~1]$, and $H(t) = 0$ for $t \in [1,~\infty)$, where $a$ is selected from $[0,~5]$.
The transmission time comparison of $p_{0.01}^{\mathrm{R}}$ ($\varepsilon = 0.01$), $p_{0.01}^{\mathrm{E}}$ ($\varepsilon = 0.01$), $p_{0.01}^{\mathrm{M}}$ ($\varepsilon = 0.01$), $p_{0.01}^{\mathrm{G}}$ ($\varepsilon = 0.01$), and $p_{0.2}^{\mathrm{R}}$ ($\varepsilon = 0.2$) is given in \figref{fig:TTO Problem Simulations T Star is Infinite}.
We can see that $\mathcal{T}^* = \mathcal{T} (p_{\varepsilon}^{\mathrm{R}}, E(0), Q(0), H_o) = \infty$ (for all $\varepsilon$).
When $\varepsilon = 0.01$, for $p_{0.01}^{\mathrm{R}}$, the region of $h$ having finite transmission time is $a \in [1.1,~5]$, but for $p_{0.01}^{\mathrm{E}}$, $p_{0.01}^{\mathrm{M}}$, or $p_{0.01}^{\mathrm{G}}$, this region is much smaller ($a \in [2.7,~5]$, $a \in [2.6,~5]$ and $a \in [2.9,~5]$, respectively).
Hence, if the actual energy-harvesting rate is with $a = 2.5$, the robust-optimal design $p_{0.01}^{\mathrm{R}}$ results in a transmission time of $1.33$, while $p_{0.01}^{\mathrm{E}}$, $p_{0.01}^{\mathrm{M}}$ and $p_{0.01}^{\mathrm{G}}$ return an infinite transmission time.
Additionally, since $0.2$ is a multiple of $0.01$, \propref{prop:The Inclusion Property of Higher Resolution} tells that $\mathcal{H}_f(p_{0.01}^{\mathrm{R}}, E(0), Q(0)) \supset \mathcal{H}_f(p_{0.2}^{\mathrm{R}}, E(0), Q(0))$, which is also verified in \figref{fig:TTO Problem Simulations T Star is Infinite}.

\begin{figure}[h]
\centering
\subfigure[]{\includegraphics [width=0.7\columnwidth]{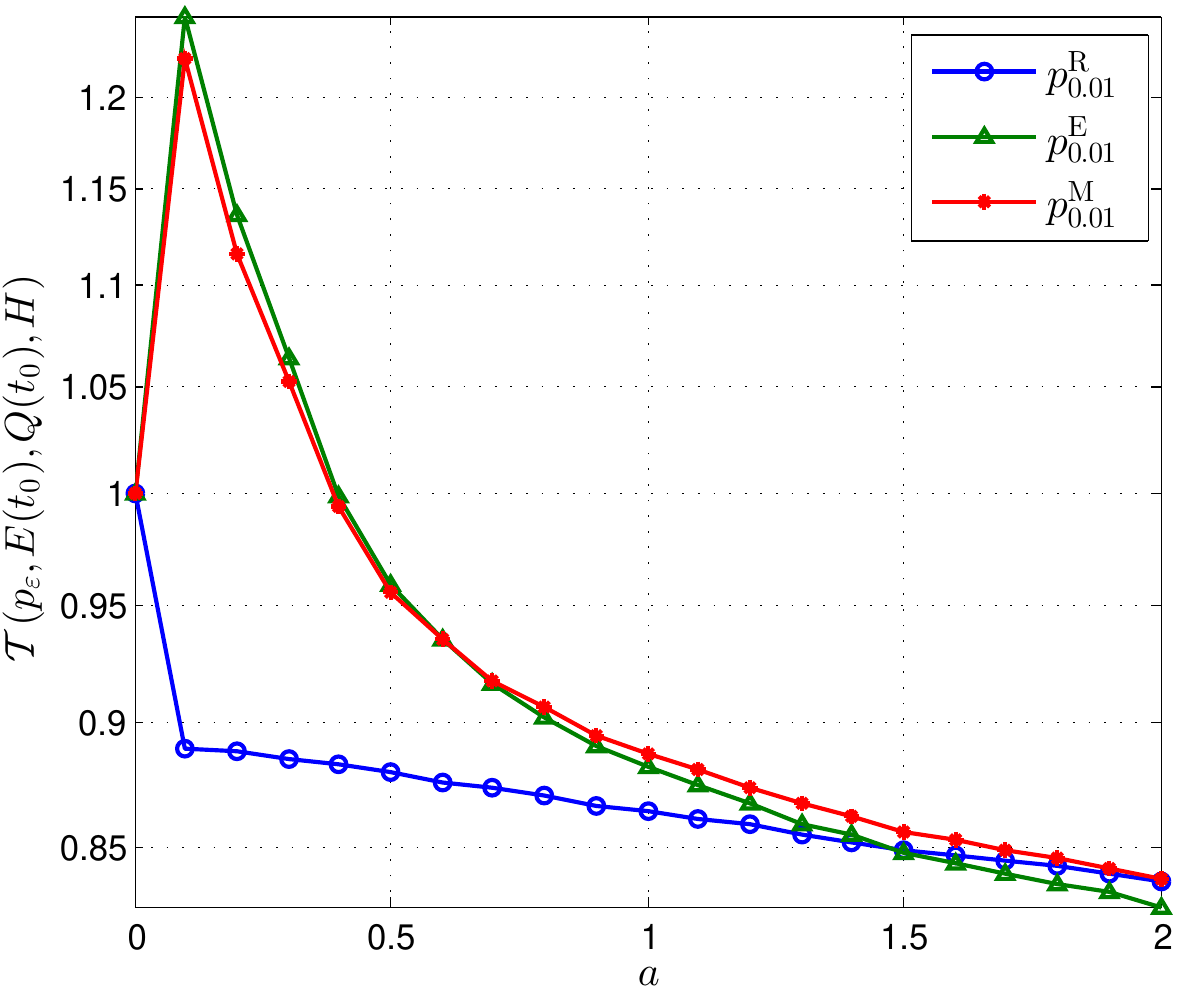}\label{fig:Simulation Scenario 3a}}\\
\subfigure[]{\includegraphics [width=0.7\columnwidth]{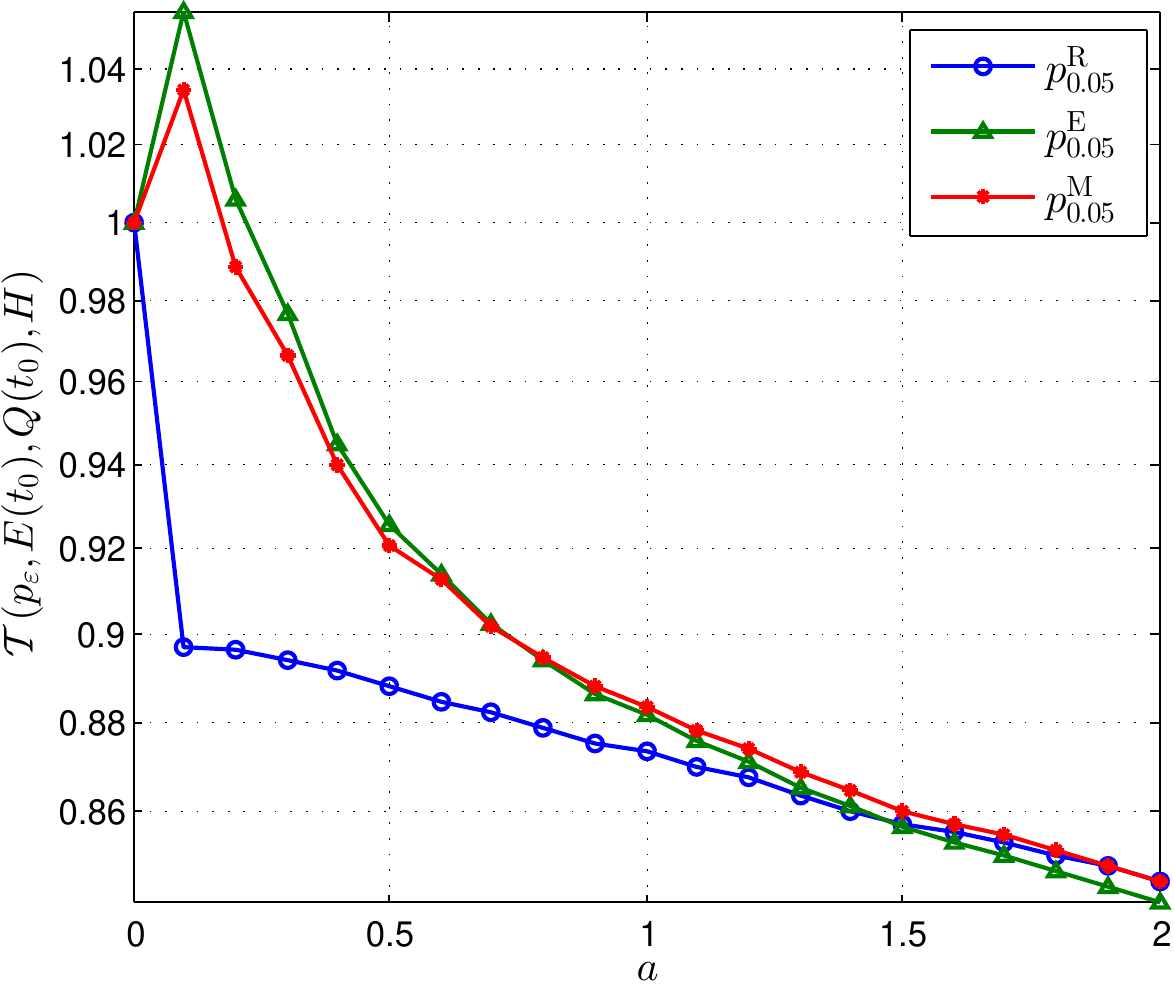}\label{fig:Simulation Scenario 3b}}

\caption{Transmission time comparisons:
(a) $\varepsilon = 0.01$;
(b) $\varepsilon = 0.05$.
}\label{fig:Simulation Scenario 3}
\end{figure}

\textbf{Scenario 3.}
We set the initial battery energy as $E(0) = 1$, the initial data queue length as $Q(0) = 1$, and the maximum transmit power as $p_{\max} = 3$.
The energy-harvesting rate is chosen as a modified compound Poisson process $H(t) = \big[\sum_{i = 1}^{N(t)} D_i(t)\big]^+$, where $\{N(t)\colon t \geq 0\}$ is a Poisson process with rate $\lambda = 2$, and $D_i(t)$ is a Gaussian random variable with mean $a|\sin t|$ ($a \in [0,~5]$) and variance $1$.
The transmission time comparisons of $p_{\varepsilon}^{\mathrm{R}}$, $p_{\varepsilon}^{\mathrm{E}}$, and $p_{\varepsilon}^{\mathrm{M}}$ for $\varepsilon = 0.01,0.05$ are given in \figref{fig:Simulation Scenario 3}.
The performance of the greedy algorithm is not included as it is much worse than all other algorithms.
In \figref{fig:Simulation Scenario 3a}, the worst-case transmission times of $p_{0.01}^{\mathrm{E}}$ and $p_{0.01}^{\mathrm{M}}$ are $24.6\%$ and $22.1\%$ larger than that of $p_{0.01}^{\mathrm{R}}$, respectively.
Similarly, in \figref{fig:Simulation Scenario 3b}, the worst-case transmission time of $p_{0.05}^{\mathrm{R}}$ is smaller than those of $p_{0.05}^{\mathrm{E}}$ and $p_{0.05}^{\mathrm{M}}$.
Comparing \figref{fig:Simulation Scenario 3a} with \figref{fig:Simulation Scenario 3b}, we can see that a larger triggering threshold actually reduces the worst-case transmission time for the estimation based algorithms $p_{\varepsilon}^{\mathrm{E}}$ and $p_{\varepsilon}^{\mathrm{M}}$.
This is because $H(t)$ is highly non-stationary and a larger triggering threshold is less sensitive to the rapid and non-stationary fluctuation in $H(t)$ which results in a better worst-case performance for the estimation-based algorithms.
For our robust-optimal algorithm $p_{\varepsilon}^{\mathrm{R}}$, the worst-case transmission times are the same for $\varepsilon = 0.01$ and $\varepsilon = 0.05$, which coincides with our conclusion in \secref{sec:Discussions on Triggering Condition}.

\section{Conclusion}\label{sec:Conclusion}

We have solved the transmission-time minimization problems for an energy harvesting transmitter, where the future energy-harvesting rate is totally unknown.
Specifically, our design is based on two advanced methods in cybernetics:
\begin{itemize}
\item   Event-trigger control:
        The Event Detector (ED) continuously monitors the battery energy and triggers a new event when it experiences some significant change from the last event.
        Whenever an event is trigger, the Transmission Planner (TP) uses the current information of the battery energy and the data queue to update the transmit power based on robust-optimal control.
        The event-trigger control framework is summarized in \algref{alg:Event-Trigger Based Control}.
\item   Robust-optimal control:
        It minimizes the worst-case transmission time such that the actual transmission time is guaranteed to be below this level, no matter what energy-harvesting rate is imposed.
        If the worst-case transmission time is always infinite for all possible transmit power design, then the robust-optimal control guarantees the largest set of energy-harvesting rate to have a finite actual transmission time.
        The robust-optimal transmit power design is given in \thmref{thm:Optimal Solution to RTT Problem}.
\end{itemize}
For future work, one can adopt the approach used in this work to design transmission protocols for other objectives, such as throughput maximization, with no knowledge on the future behavior of energy-harvesting rate and data arrival process.
Additionally, it is interesting to see how the robust-optimal solution performs using experimental measurements (e.g.,~\cite{GuendaL2014}) of the energy-harvesting rate.

\appendices
%
%


\section{Proof of \lemref{lem:Criterion on Points in Reachable Set}}\label{apx:Proof of Lemma 1}

Necessity.~If~\eqref{eqn:Points in Reachable Set Other Than starting point} holds, there exists a $t \geq t_n$ such that $\mathfrak{R}_n$ is not empty, and we have
\begin{align}\label{eqn:Division of DQ and DE}
\frac {Q(t_n) - \widetilde{Q}_n} {E(t_n) - \widetilde{E}_n} = \frac {\int_{t_n}^{t}r(\tilde{p}_{n,\varepsilon}(\tau))\mathrm{d}\tau} {\int_{t_n}^{t}\tilde{p}_{n,\varepsilon}(\tau)\mathrm{d}\tau}.
\end{align}
According to~\eqref{eqn:Rate-Power Line}, equation~\eqref{eqn:Division of DQ and DE} can be further rewritten as
\begin{align}\label{eqn:Balance Equation}
\int_{t_n}^{t} \left[K_n \tilde{p}_{n,\varepsilon}(\tau) - r(\tilde{p}_{n,\varepsilon}(\tau))\right] \mathrm{d}\tau = 0.
\end{align}
If no RPE exists, then either $K_n \tilde{p}_{n,\varepsilon}(\tau) > r(\tilde{p}_{n,\varepsilon}(\tau))$ or $K_n \tilde{p}_{n,\varepsilon}(\tau) < r(\tilde{p}_{n,\varepsilon}(\tau))$ holds for $[t_n,~t_n+T_n]$, which contradicts with~\eqref{eqn:Balance Equation}.
Therefore, the RPE exists.

Sufficiency.~If the RPE exists, we set $\tilde{p}_{n,\varepsilon}(t) = p_n^e$ as the transmit power.
Note that $K_n p_n^e = r(p_n^e)$, and hence~\eqref{eqn:Points in Reachable Set Other Than starting point} holds.

\section{Proof of \propref{prop:Expressions of Reachable Set}}\label{apx:Proof of Proposition 1}

For the simplicity of this proof, we label $A_1 = \mathfrak{R}_n\backslash\{(E(t_n),Q(t_n))\}$ and $A_2 = \{(\widetilde{E}_n,\widetilde{Q}_n): K_{\min} \leq K_n < K_{\max}, 0 \leq \widetilde{E}_n < E(t_n), 0 \leq \widetilde{Q}_n < Q(t_n)\}$.

i)~$A_1 \subseteq A_2$:
$\forall (\widetilde{E}_n(t),\widetilde{Q}_n(t)) \in A_1$, the RPE exists according to \lemref{lem:Criterion on Points in Reachable Set}.
Hence, $K_n = r(p_n^e) / p_n^e$.
Since $r(p)$ is strictly concave for $p$ and $r(0) = 0$,
\begin{align}\label{eqn:Concave Property of Rate Function}
K_n = \frac {r(p_n^e)} {p_n^e} = \frac {r(\overline{p}_e) - r(0)} {\overline{p}_e - 0}
\end{align}
is strictly decreasing in $(0,~p_{\max}]$.
Therefore, $K_{\min} \leq K_n < K_{\max}$, and $(\widetilde{E}_n,\widetilde{Q}_n) \in A_2$.

ii)~$A_2 \subseteq A_1$:
$\forall (\widetilde{E}_n,\widetilde{Q}_n) \in A_2$, the RPE exists according to~\eqref{eqn:Concave Property of Rate Function}.
Thus,~$(\widetilde{E}_n,\widetilde{Q}_n) \in A_1$.

To sum up, $A_1 = A_2$ and~\eqref{eqn:Expressions of Reachable Set} holds.

\section{Proof of \propref{prop:Time-Optimal PTP}}\label{apx:Proof of Proposition 2}

Since $r(\tilde{p}_{n,\varepsilon}(t)) = \log_2(1 + \tilde{p}_{n,\varepsilon}(t))$ is strictly increasing, concave and non-negative, $\forall \tilde{p}_{n,\varepsilon}(t) \in [0,~p_{\max}]$, $r(\tilde{p}_{n,\varepsilon}(t))$ can be expressed by
\begin{align}\label{eqn:Expression for r}
r(\tilde{p}_{n,\varepsilon}(t)) = K'_e \tilde{p}_{n,\varepsilon}(t) + b - \epsilon(\tilde{p}_{n,\varepsilon}(t)),
\end{align}
where $b$ is a positive constant, and $K'_e$ is the derivative of $r(\tilde{p}_{n,\varepsilon}(t))$ at $\tilde{p}_{n,\varepsilon}(t) = p_n^e$, which implies $K'_e := \lim_{\tilde{p}_{n,\varepsilon}(t)\rightarrow p_n^e} {\mathrm{d} r(\tilde{p}_{n,\varepsilon}(t))} / {\mathrm{d} \tilde{p}_{n,\varepsilon}(t)}$.
In~\eqref{eqn:Expression for r}, $\epsilon(\tilde{p}_{n,\varepsilon}(t)) \geq 0$, and $\epsilon(\tilde{p}_{n,\varepsilon}(t)) = 0$ holds only when $\tilde{p}_{n,\varepsilon}(t) = p_n^e$.

Recall the proof of necessity in \lemref{lem:Criterion on Points in Reachable Set} that if~\eqref{eqn:Points in Reachable Set Other Than starting point} holds, then~\eqref{eqn:Balance Equation} is satisfied.
We break~\eqref{eqn:Balance Equation} down into 3 parts by Lebesgue integral,
\begin{align}\label{eqn:Balance Equation into 3 Parts}
\int_{S_1} \Lambda(t) \mathrm{d} t
+ \int_{S_2} \Lambda(t) \mathrm{d} t
+ \int_{S_3} \Lambda(t) \mathrm{d} t = 0,
\end{align}
where $\Lambda(t) = \left[r(\tilde{p}_{n,\varepsilon}(t)) - K_n \tilde{p}_{n,\varepsilon}(t)\right]$, $S_1 = \{t:\tilde{p}_{n,\varepsilon}(t) < p_n^e\}$, $S_2 = \{t:\tilde{p}_{n,\varepsilon}(t) > p_n^e\}$ and $S_3 = \{t:\tilde{p}_{n,\varepsilon}(t) = p_n^e\}$.

Let $\overline{p}$ be the average transmit power over $[t_n,~t_n + T_n]$, and we have
\begin{align}\label{eqn:Pbar}
\begin{split}
\overline{p} &\stackrel {(a)} {=} \frac {\int_{S_1} \tilde{p}_{n,\varepsilon}(t) \mathrm{d} t + \int_{S_2} \tilde{p}_{n,\varepsilon}(t) \mathrm{d} t + \int_{S_3} p_n^e \mathrm{d} t} {\mu \left(S_1\right) + \mu \left(S_2\right) + \mu \left(S_3\right)}\\
&\stackrel {(b)} {=} \frac {\int_{S_1} r\left(\tilde{p}_{n,\varepsilon}(t)\right) \mathrm{d} t + \int_{S_2} r\left(\tilde{p}_{n,\varepsilon}(t)\right) \mathrm{d} t + \int_{S_3} r\left(p_n^e\right) \mathrm{d} t} {K_n\left[\mu \left(S_1\right) + \mu \left(S_2\right) + \mu \left(S_3\right)\right]}\\
&\stackrel {(c)} {=} \frac {K'_e} {K_n} \overline{p} + \frac {b} {K_n} - \Delta,
\end{split}
\end{align}
where
\begin{align}\label{eqn:Delta}
\Delta = \frac {\int_{S_1} \epsilon\left(\tilde{p}_{n,\varepsilon}(t)\right) \mathrm{d} t + \int_{S_2} \epsilon\left(\tilde{p}_{n,\varepsilon}(t)\right) \mathrm{d} t} {K_n\left[\mu \left(S_1\right) + \mu \left(S_2\right) + \mu \left(S_3\right)\right]} \geq 0.
\end{align}
In~\eqref{eqn:Pbar}, (a) represents the average power, and (b) is from~\eqref{eqn:Balance Equation into 3 Parts}, and (c) is derived by employing~\eqref{eqn:Expression for r}.
In~\eqref{eqn:Delta}, the equality holds only when $\mu \left(S_1\right) = 0$ and $\mu \left(S_2\right) = 0$, i.e.,~\eqref{eqn:Time-Optimal PTP} holds\footnote{Mathematically, formula~\eqref{eqn:Time-Optimal PTP} holds almost everywhere (a.e.) in the duration $[t_n,~t_n+T_n]$ except some sub-durations with zero measures.
Practically, since the transmit power cannot change that fast, we exclude the zero-measure cases and the term a.e. is omitted.}.
From~\eqref{eqn:Pbar}, the explicit form for $\overline{p}$ is
\begin{align}
\overline{p} = \frac {b - K_n \Delta} {K_n - K'_e} \leq \frac {b} {K_n - K'_e} = p_n^e.
\end{align}
Therefore,
\begin{align}\label{eqn:T_n and Optimal T_n}
\begin{split}
T_n &= \mu \left(S_1\right) + \mu \left(S_2\right) + \mu \left(S_3\right)\\
&= \frac {E(t_n) - \widetilde{E}_n} {\overline{p}} \geq \frac {E(t_n) - \widetilde{E}_n} {p_n^e} = \underline{T}_n,
\end{split}
\end{align}
where the equality holds if and only if~\eqref{eqn:Time-Optimal PTP} holds according to~\eqref{eqn:Delta}.
Thus,~\eqref{eqn:Time-Optimal PTP} is the time-optimal transmission power.
The uniqueness for~\eqref{eqn:Time-Optimal PTP} is obvious, otherwise~\eqref{eqn:T_n and Optimal T_n} cannot hold.

Note that $K_n p_n^e = r(p_n^e)$, and hence~\eqref{eqn:Optimal Planned Transimission Time} is obtained.


\section{Proof of \corref{cor:Minimum Transmission-Time to Clear the Data Queue from the TP Side}}\label{apx:Proof of Corollary 1}

Taking the partial derivative of $\underline{T}_n\big[\widetilde{E}_n,\widetilde{Q}_n\big]$ in~\eqref{eqn:Optimal Planned Transimission Time} w.r.t. $\widetilde{E}_n$, we have
\begin{align}\label{eqn:Partial Derivative of Planned Transmission Time w.r.t. E}
\begin{split}
\frac{\partial \underline{T}_n\big[\widetilde{E}_n,\widetilde{Q}_n\big]}{\partial \widetilde{E}_n} &= \frac{\partial}{\partial \widetilde{E}_n}\left[\frac {E(t_n) - \widetilde{E}_n}{p_n^e}\right]\\
&= - \frac{p_n^e + \frac{\partial p_n^e}{\partial \widetilde{E}_n}\left[E(t_n) - \widetilde{E}_n\right]}{(p_n^e)^2}.
\end{split}
\end{align}
Employing the chain rule, we have
\begin{align}\label{eqn:Chain Rule on Partial p Partial E}
\frac{\partial p_n^e}{\partial \widetilde{E}_n} = \frac{\partial p_n^e}{\partial K_n} \frac{\partial K_n}{\partial \widetilde{E}_n} = \frac{p_n^e}{K'_n - K_n} \frac{K_n}{E(t_n) - \widetilde{E}_n},
\end{align}
where ${\partial p_n^e}/{\partial K_n}$ is derived by using the implicit differentiation on $K_n p_n^e = \log_2(1 + p_n^e)$, and ${\partial K_n}/{\partial \widetilde{E}_n}$ is obtained by~\eqref{eqn:Rate-Power Line}.
In~\eqref{eqn:Chain Rule on Partial p Partial E}, $K'_n$ is the derivative of $\log_2(1 + p_n^e)$, i.e., $K'_n := 1/[\ln 2 (1 + p_n^e)]$.
Since $K_n$ and $K'_n$ can be rewritten as
\begin{align}\label{eqn:Kn and K'n Rewritten}
K_n = \frac{\log_2 (1 + p_n^e) - 0}{p_n^e - 0}, \quad K'_n = \lim_{p \to p_n^e}\frac{\log_2 (1 + p_n^e) - p}{p_n^e - p},
\end{align}
we know $K_n > K'_n$ due to $p_n^e > p > 0$.
By~\eqref{eqn:Chain Rule on Partial p Partial E}, equation~\eqref{eqn:Partial Derivative of Planned Transmission Time w.r.t. E} can be rewritten as
\begin{align}\label{eqn:Partial Derivative of Planned Transmission Time w.r.t. E Rewritten}
\frac{\partial \underline{T}_n\big[\widetilde{E}_n,\widetilde{Q}_n\big]}{\partial \widetilde{E}_n} = \frac{K'_n}{p_n^e(K_n - K'_n)} \stackrel {(a)}{>} 0,
\end{align}
where $(a)$ follows from $K_n > K'_n$.
Thus, with fixed $\widetilde{Q}_n$, the planned transmission time $\underline{T}_n\big[\widetilde{E}_n,\widetilde{Q}_n\big]$ increases with $\widetilde{E}_n$.
This means that:
If $K_n^{\mathrm{bal}} < K_{\min}$, the minimum $\underline{T}_n\big[\widetilde{E}_n,0\big]$ achieves at point $b$ in \figref{fig:Reachable Set Battery Energy Surplus}, and we can calculate the battery energy of point $b$ as $E(t_n) - {Q(t_n)}/{K_{\min}}$ by the equation of $L_1$ (see the caption in \figref{fig:Reachable Sets}), which implies~\eqref{eqnincor: Energy-Abundant Case} holds.
Since $L_1$ in \figref{fig:Reachable Set Battery Energy Surplus} is with slope $K_{\min}$, from~\figref{fig:Rate-power line, rate function, RPE and IRPE} we have $p_n^e = p_{\max}$.
Similarly, if $K_{\min} \leq K_n^{\mathrm{bal}} < K_{\max}$, we can derive~\eqref{eqnincor: Energy-Balanced Case} and~\eqref{eqn:pbaln}.

For $K_n^{\mathrm{bal}} \geq K_{\max}$, we cannot find any end point with $\widetilde{Q}_n = 0$, according to \figref{fig:Reachable Set Data Queue Surplus}.

\section{Proof of \thmref{thm:Optimal Solution to RTT Problem}}\label{apx:Proof of Theorem 1}

Before starting, we give two lemmas: \lemref{lem:The Worst-Case Energy-Harvesting Rate of Designed PTP} tells that the worst-case energy-harvesting rate of the designed PTP $\tilde{p}_{n,\varepsilon}^{\mathrm{R}}$ is $H_o: t \mapsto 0$ ($t \in [t_0,~\infty)$) if the reachable set $\mathfrak{R}_0$ is not in energy-scarce case; \lemref{lem:Necessary and Sufficient Condition for T < infty} indicates when $\mathcal{T}^*$ is finite.

\begin{lemma}\label{lem:The Worst-Case Energy-Harvesting Rate of Designed PTP}
If the initial battery energy $E(t_0)$ and initial data queue $Q(t_0)$ satisfies $Q(t_0)/E(t_0) = K_0^\mathrm{bal} < K_{\max}$, then the following holds
\begin{align}\label{eqn:The Worst-Case Energy-Harvesting Rate of Designed PTP}
\sup_{H \in \mathcal{H}}\mathcal{T}(p_{\varepsilon}^{\mathrm{R}}, E(t_0), Q(t_0), H) = \mathcal{T}(p_{\varepsilon}^{\mathrm{R}}, E(t_0), Q(t_0), H_o) < \infty.
\end{align}
\end{lemma}

\begin{IEEEproof}
Firstly, it can be easily obtained that
\begin{align}\label{eqninpf:The Worst-Case Energy-Harvesting Rate of Designed PTP 1}
\sup_{H \in \mathcal{H}}\mathcal{T}(p_{\varepsilon}^{\mathrm{R}}, E(t_0), Q(t_0), H) \geq \mathcal{T}(p_{\varepsilon}^{\mathrm{R}}, E(t_0), Q(t_0), H_o).
\end{align}
Secondly, for energy-harvesting rate $H_o$, there are no triggered events, since no energy comes in $[t_0,~\infty)$.
For $H \neq H_o$, there should be $k \in \overline{\mathbb{Z}}_+$ triggered events.
If $k = 0$, then the transmission time will be the same as that for $H_o$.
If $k > 0$, then in the first triggered event, then $E(t_1)$ is
\begin{align}\label{eqninpf:The Worst-Case Energy-Harvesting Rate of Designed PTP 2}
\!\!\!\!\begin{cases}
E(t_0) - (t_1 - t_0) p_{\max} + \varepsilon, & K_0^{\mathrm{bal}} < K_{\min},\\
E(t_0) - (t_1 - t_0) p_0^{\mathrm{bal}} + \varepsilon, & K_{\min} \leq K_0^{\mathrm{bal}} < K_{\max}.
\end{cases}
\end{align}
If such calculated $E(t_1) \leq 0$, then the transmission time is still the same as that for $H_o$.
If $E(t_1) > 0$, then $K_1^{\mathrm{bal}}$ is
\begin{align}\label{eqninpf:The Worst-Case Energy-Harvesting Rate of Designed PTP 3}
\begin{cases}
\frac {Q(t_0) - (t_1 - t_0) r(p_{\max})}{E(t_0) - (t_1 - t_0) p_{\max} + \varepsilon} \leq K_0^{\mathrm{bal}}, & K_0^{\mathrm{bal}} < K_{\min},\\
\frac {Q(t_0) - (t_1 - t_0) r(p_0^{\mathrm{bal}})}{E(t_0) - (t_1 - t_0) p_0^{\mathrm{bal}} + \varepsilon} < K_0^{\mathrm{bal}}, & K_{\min} \leq K_0^{\mathrm{bal}} < K_{\max}.
\end{cases}
\end{align}
According to \figref{fig:Rate-power line, rate function, RPE and IRPE} that $p_n^e$ decreases w.r.t. $K_n$, we have $p_0^{\mathrm{bal}} \leq p_1^{\mathrm{bal}}$.
Therefore, the transmission time becomes smaller.
To sum up, $\forall H \in \mathcal{H}$ the following inequality holds
\begin{align}\label{eqninpf:The Worst-Case Energy-Harvesting Rate of Designed PTP 4}
\mathcal{T}(p_{\varepsilon}^{\mathrm{R}}, E(t_0), Q(t_0), H) \leq \mathcal{T}(p_{\varepsilon}^{\mathrm{R}}, E(t_0), Q(t_0), H_o),
\end{align}
which means
\begin{align}\label{eqninpf:The Worst-Case Energy-Harvesting Rate of Designed PTP 5}
\sup_{H \in \mathcal{H}}\mathcal{T}(p_{\varepsilon}^{\mathrm{R}}, E(t_0), Q(t_0), H) \leq \mathcal{T}(p_{\varepsilon}^{\mathrm{R}}, E(t_0), Q(t_0), H_o).
\end{align}
Now, combining inequality~\eqref{eqninpf:The Worst-Case Energy-Harvesting Rate of Designed PTP 5} with inequality~\eqref{eqninpf:The Worst-Case Energy-Harvesting Rate of Designed PTP 1} and noticing that $\mathcal{T}(p_{\varepsilon}^{\mathrm{R}}, E(t_0), Q(t_0), H_o) < \infty$ (since the corresponding reachable set for $n = 0$ is energy-abundant or energy-balanced), we can finally derive~\eqref{eqn:The Worst-Case Energy-Harvesting Rate of Designed PTP}.
\end{IEEEproof}

\begin{lemma}\label{lem:Necessary and Sufficient Condition for T < infty}
$\mathcal{T}^* < \infty$ if and only if $Q(t_0)/E(t_0) = K_0^\mathrm{bal} < K_{\max}$.
\end{lemma}

\begin{IEEEproof}
Necessity.~By contrapositive, we should prove that: if $K_0^{\mathrm{bal}} \geq K_{\max}$, then $\mathcal{T}^* = \infty$.
Under $H_o : t \mapsto 0$ ($t \in [t_0,~\infty)$), we have $\mathcal{T}(p_{\varepsilon}, E(t_0), Q(t_0), H_o) = \infty$ for any $p_{\varepsilon} \in \mathfrak{P}_{\varepsilon}$, which implies
\begin{align}\label{eqninpf:Necessary and Sufficient Condition for T < infty 1}
\begin{split}
\mathcal{T}^* &= \inf_{p_{\varepsilon} \in \mathfrak{P}_{\varepsilon}} \sup_{H \in \mathcal{H}} \mathcal{T}(p_{\varepsilon}, E(t_0), Q(t_0), H)\\
&\geq \inf_{p_{\varepsilon} \in \mathfrak{P}_{\varepsilon}} \mathcal{T}(p_{\varepsilon}, E(t_0), Q(t_0), H_o) = \infty.
\end{split}
\end{align}
Therefore, $\mathcal{T}^* = \infty$.

Sufficiency.~We prove that if $K_0^\mathrm{bal} < K_{\max}$, then $\mathcal{T}^* < \infty$.
Firstly, we have
\begin{align}\label{eqninpf:Necessary and Sufficient Condition for T < infty 2}
\begin{split}
\mathcal{T}^* &= \inf_{p_{\varepsilon} \in \mathfrak{P}_{\varepsilon}} \sup_{H \in \mathcal{H}} \mathcal{T}(p_{\varepsilon}, E(t_0), Q(t_0), H)\\
&\leq \sup_{H \in \mathcal{H}} \mathcal{T}(p_{\varepsilon}^{\mathrm{R}}, E(t_0), Q(t_0), H),
\end{split}
\end{align}
which combined with $K_0^\mathrm{bal} < K_{\max}$ and \lemref{lem:The Worst-Case Energy-Harvesting Rate of Designed PTP} implies that
\begin{align}\label{eqninpf:Necessary and Sufficient Condition for T < infty 3}
\begin{split}
\mathcal{T}^* &\leq \sup_{H \in \mathcal{H}} \mathcal{T}(p_{\varepsilon}^{\mathrm{R}}, E(t_0), Q(t_0), H)\\
&\leq \mathcal{T}(p_{\varepsilon}^{\mathrm{R}}, E(t_0), Q(t_0), H_o) < \infty.
\end{split}
\end{align}
\end{IEEEproof}

Now, we start the proof of \thmref{thm:Optimal Solution to RTT Problem}.

We divide this proof into two parts corresponding to $\mathcal{T}^* < \infty$ (see \subpref{subp:Transmission-Time-Minimization Problem}) and $\mathcal{T}^* = \infty$ (see \subpref{subp:Maximum-Energy-Set Problem}), respectively.

i) For $\mathcal{T}^* < \infty$, from \lemref{lem:Necessary and Sufficient Condition for T < infty}, we know $K_0^{\mathrm{bal}} < K_{\max}$ holds.

Let the optimal solution be $\tilde{p}_{n,\varepsilon}^*$ (implementing $p_{\varepsilon}^*$) and the corresponding worst-case energy-harvesting rate be $H_{p_{\varepsilon}^*}$, i.e., $\mathcal{T}^* = \mathcal{T}(p_{\varepsilon}^*,E(t_0), Q(t_0),H_{p_{\varepsilon}^*})$.
We have
\begin{multline}\label{eqninpf:Robust Optimal TTOPTP 1}
\mathcal{T}(p_{\varepsilon}^*,E(t_0), Q(t_0),H_{p_{\varepsilon}^*}) = \!\mathcal{T}^* \\\leq \sup_{H \in \mathcal{H}} \mathcal{T}(p_{\varepsilon}^{\mathrm{R}},E(t_0), Q(t_0),H) \stackrel {(a)}{=} \mathcal{T}(p_{\varepsilon}^{\mathrm{R}},E(t_0), Q(t_0),H_o),
\end{multline}
where $(a)$ follows from \lemref{lem:The Worst-Case Energy-Harvesting Rate of Designed PTP}, since for $K_0^{\mathrm{bal}} < K_{\max}$, the worst-case energy-harvesting rate of $p_{\varepsilon}^\mathrm{R}$ is $H_o$.

On the other hand, we can derive
\begin{align}\label{eqninpf:Robust Optimal TTOPTP 2}
\begin{split}
\mathcal{T}(p_{\varepsilon}^*,E(t_0), Q(t_0),H_{p^*}) &\geq \mathcal{T}(p_{\varepsilon}^*,E(t_0), Q(t_0),H_o)\\
&\stackrel {(b)}{\geq} \mathcal{T}(p_{\varepsilon}^{\mathrm{R}},E(t_0), Q(t_0),H_o),
\end{split}
\end{align}
where $(b)$ follows from the fact that $p_{\varepsilon}^{\mathrm{R}}$ is time optimal under $H_o$.
This is because, in this case, event $1$ is never triggered, and the PTP is always equal to the actual transmit power [see~\eqref{eqn:Relationship between Transmit Power and PTP}], which means that: for $K_0^{\mathrm{bal}} < K_{\max}$, the actual transmit power is optimal according to \corref{cor:Minimum Transmission-Time to Clear the Data Queue from the TP Side}.
Therefore, combining~\eqref{eqninpf:Robust Optimal TTOPTP 1} and~\eqref{eqninpf:Robust Optimal TTOPTP 2}, we have $p_{\varepsilon}^* = p_{\varepsilon}^{\mathrm{R}}$, i.e., $\tilde{p}_{n,\varepsilon}^* = \tilde{p}_{n,\varepsilon}^\mathrm{R}$.

ii) For $\mathcal{T}^* = \infty$, from \lemref{lem:Necessary and Sufficient Condition for T < infty} we know $K_0^{\mathrm{bal}} \geq K_{\max}$ holds.

We will show $p_{\varepsilon}^{\mathrm{R}}$ satisfies~\eqref{eqn:Maximum-Energy-Set Problem} in \subpref{subp:Maximum-Energy-Set Problem} by
\begin{align}\label{eqninpf:Robust Optimal TTOPTP 3}
\mathcal{H}_f(p_{\varepsilon}, E(t_0), Q(t_0)) \subseteq \mathcal{H}_f(p_{\varepsilon}^{\mathrm{R}}, E(t_0), Q(t_0)), \quad \forall p_{\varepsilon} \in \mathfrak{P}_{\varepsilon},
\end{align}
or equivalently, for all $p_{\varepsilon} \in \mathfrak{P}_{\varepsilon}$, $\forall H \in \mathcal{H}_f(p_{\varepsilon}, E(t_0), Q(t_0))$, $H$ should also be in the set $\mathcal{H}_f(p_{\varepsilon}^{\mathrm{R}}, E(t_0), Q(t_0))$.

Now, $\forall p_{\varepsilon} \in \mathfrak{P}_{\varepsilon}$, and $\forall H \in \mathcal{H}_f(p_{\varepsilon}, E(t_0), Q(t_0))$, we have
\begin{align}\label{eqninpf:Data Queue Equation}
\int_{t_0}^{t_0+\mathcal{T}} \log_2 (1 + p_{\varepsilon}(\tau)) \mathrm{d} \tau = Q(t_0),
\end{align}
where $\mathcal{T} = \mathcal{T}(p_{\varepsilon}, E(t_0), Q(t_0), H) < \infty$.
Since function $H$ is Lebesgue integrable over any subset of $\overline{\mathbb{R}}_+$ with finite measure, the integral $\int_{t_0}^{t_0+\mathcal{T}} H(\tau) \mathrm{d}\tau$ is finite, which means the total number of triggered events is also finite.
Assuming event $N$ is the last triggered event, we have
\begin{align}\label{eqninpf:Energy Equation}
\int_{t_0}^{t_0+\mathcal{T}} p_{\varepsilon}(\tau) \mathrm{d} \tau = E(t_0) + N \varepsilon,
\end{align}
in which the value $\varepsilon$ is the harvested energy in each event, since the integral of $H(t)$ is continuous and the event is triggered when $\int_{t_{n-1}}^{t_n} H(\tau) \mathrm{d}\tau = \varepsilon$.
Thus, $N \varepsilon$ is the total amount of harvested energy from $t_0$ to $t_0 + \mathcal{T}$ that can be seen from the TP side.

If we assume the amount of battery energy $E(t_0) + N \varepsilon$ is available at $t_0$, then $p_{\varepsilon}$ is still be a valid transmit power design to clear the data queue $Q(t_0)$, and the corresponding transmission time is also the same, i.e., $\mathcal{T} = \mathcal{T}(p_{\varepsilon}, E(t_0) + N \varepsilon, Q(t_0), H_o)$.\footnote{After moving the future energy to $t_0$, the transmission time should be $\mathcal{T}(p_{\varepsilon}, E(t_0) + N \varepsilon, Q(t_0), H_{\mathrm{left}})$ where $H_{\mathrm{left}}$ is an equivalent energy-harvesting rate but not necessarily $H_o$.
However, since $H_{\mathrm{left}}$ cannot trigger any event, we have $\mathcal{T}(p_{\varepsilon}, E(t_0) + N \varepsilon, Q(t_0), H_{\mathrm{left}}) = \mathcal{T}(p_{\varepsilon}, E(t_0) + N \varepsilon, Q(t_0), H_o)$}.
With \corref{cor:Minimum Transmission-Time to Clear the Data Queue from the TP Side}, we know that $p_{\varepsilon}^{\mathrm{R}}$ is time-optimal, i.e., $\mathcal{T}(p_{\varepsilon}^{\mathrm{R}}, E(t_0) + N \varepsilon, Q(t_0), H_o) \leq \mathcal{T}(p_{\varepsilon}, E(t_0) + N \varepsilon, Q(t_0), H_o) < \infty$, which implies
\begin{align}\label{eqninpf:Robust Optimal TTOPTP 4}
\frac {Q(t_0)} {E(t_0) + N \varepsilon} \leq K_{\max}.
\end{align}

Then, we prove $\mathcal{T}(p_{\varepsilon}^{\mathrm{R}}, E(t_0), Q(t_0), H)$ is finite.
In event $0$:
\begin{itemize}
\item[1)]   If $K_0^{\mathrm{bal}} = Q(t_0)/E(t_0) \leq K_{\max}$, then the transmission time is finite, because even if no energy arrives in $[t_0,~\infty)$, the transmitter can still clean up the data queue by~\eqref{eqn:Optimal Solution of RTT Problem}.
\item[2)]   If $K_0^{\mathrm{bal}} = Q(t_0)/E(t_0) \geq K_{\max}$, then from~\eqref{eqn:Optimal Solution of RTT Problem}, the transmit power is $0$ during $[t_0,~t_1)$.
\end{itemize}
If case~2) happens, then in event $1$, we have $K_1^{\mathrm{bal}} = Q(t_0)/(E(t_0) + \varepsilon)$.
Similar to event $0$, if $K_1^{\mathrm{bal}} \leq K_{\max}$, then the transmission time is finite.
Otherwise, the transmit power is $0$ during $[t_1,~t_2)$, which makes $K_2^{\mathrm{bal}} = Q(t_0)/(E(t_0) + 2\varepsilon)$.
Proceeding forward, if case~1) happens, then the transmission time is finite, and if case~2) happens, the transmitter transmit nothing.
Here, we can always find an event $n \leq N$ such that $K_n^{\mathrm{bal}} = Q(t_0)/(E(t_0) + n\varepsilon)$, if $K_{n-1}^{\mathrm{bal}} \geq K_{\max}$.
This is guaranteed by~\eqref{eqninpf:Robust Optimal TTOPTP 4}.
Now, we have $\mathcal{T}(p_{\varepsilon}^{\mathrm{R}}, E(t_0), Q(t_0), H) < \infty$, and thus $H \in \mathcal{H}_f(p_{\varepsilon}^{\mathrm{R}}, E(t_0), Q(t_0))$.

To sum up, for any $p_{\varepsilon} \in \mathfrak{P}_{\varepsilon}$, $\forall H \in \mathcal{H}_f(p_{\varepsilon}, E(t_0), Q(t_0))$, we have $H \in \mathcal{H}_f(p_{\varepsilon}^{\mathrm{R}}, E(t_0), Q(t_0))$, which implies that $\mathcal{H}_f(p_{\varepsilon}, E(t_0), Q(t_0)) \subseteq \mathcal{H}_f(p_{\varepsilon}^{\mathrm{R}}, E(t_0), Q(t_0))$ holds for all $p_{\varepsilon} \in \mathfrak{P}_{\varepsilon}$.

\section{Proof of \propref{prop:The Inclusion Property of Higher Resolution}}\label{apx:Proof of Proposition 3}

$\forall H \in \mathcal{H}_f(p_{\varepsilon^b}^{\mathrm{R}}, E(t_0), Q(t_0))$, the transmission time is $\mathcal{T}^b = \mathcal{T}(p_{\varepsilon^b}^{\mathrm{R}}, E(t_0), Q(t_0), H) < \infty$.
Then, similar to part~ii) in the proof of \thmref{thm:Optimal Solution to RTT Problem}, we label the last triggered event number as $N^b$, and by~\eqref{eqn:Event E Happens}, the following holds
\begin{align}
N^b \varepsilon^b \leq \int_{t_0}^{t_0+\mathcal{T}^b} H(\tau) \mathrm{d}\tau < (N^b + 1) \varepsilon^b.
\end{align}
Since $\mathcal{T}^b$ is finite, similar to~\eqref{eqninpf:Robust Optimal TTOPTP 4}, $Q(t_0) / (E(t_0) + N^b \varepsilon^b) < K_{\max}$ holds.
Noting that $z \varepsilon^a = \varepsilon^b$, we have $N^b \varepsilon^b =  N^a \varepsilon^a$, where $N^a = z N^b$.
Thus, the following holds
\begin{align}
\frac {Q(t_0)} {E(t_0) + N^a \varepsilon^a} = \frac {Q(t_0)} {E(t_0) + N^b \varepsilon^b} < K_{\max},
\end{align}
which implies that for $\varepsilon^a$, the transmission time $\mathcal{T}(p_{\varepsilon^a}^{\mathrm{R}}, E(t_0), Q(t_0), H)$ is finite (similar to part~ii) in the proof of \thmref{thm:Optimal Solution to RTT Problem}).
Therefore, $H \in \mathcal{H}_f(p_{\varepsilon^a}^{\mathrm{R}}, E(t_0), Q(t_0))$, and $\mathcal{H}_f(p_{\varepsilon^a}^{\mathrm{R}}, E(t_0), Q(t_0)) \supseteq \mathcal{H}_f(p_{\varepsilon^b}^{\mathrm{R}}, E(t_0), Q(t_0))$.

On the other hand, we can find a $H \in \mathcal{H}_f(p_{\varepsilon^a}^{\mathrm{R}}, E(t_0), Q(t_0))$ such that
\begin{align}
(z-1) \varepsilon^a = \int_{t_0}^{\infty} H(\tau) \mathrm{d}\tau < z \varepsilon^a = \varepsilon^b,
\end{align}
which implies $\mathcal{T}^b$ is infinite, since for $\varepsilon^b$, the first event never comes, and $K_0^{\mathrm{bal}} \geq K_{\max}$ will stay unchanged for $[t_0,\infty)$.
Hence, $\mathcal{H}_f(p_{\varepsilon^a}^{\mathrm{R}}, E(t_0), Q(t_0)) \not\supseteq \mathcal{H}_f(p_{\varepsilon^b}^{\mathrm{R}}, E(t_0), Q(t_0))$, and therefore, combining with the result in the last paragraph, we have $\mathcal{H}_f(p_{\varepsilon^a}^{\mathrm{R}}, E(t_0), Q(t_0)) \supset \mathcal{H}_f(p_{\varepsilon^b}^{\mathrm{R}}, E(t_0), Q(t_0))$.

\bibliographystyle{IEEEtran}
\bibliography{BIB}

\begin{thebibliography}{10}
\providecommand{\url}[1]{#1}
\csname url@samestyle\endcsname
\providecommand{\newblock}{\relax}
\providecommand{\bibinfo}[2]{#2}
\providecommand{\BIBentrySTDinterwordspacing}{\spaceskip=0pt\relax}
\providecommand{\BIBentryALTinterwordstretchfactor}{4}
\providecommand{\BIBentryALTinterwordspacing}{\spaceskip=\fontdimen2\font plus
\BIBentryALTinterwordstretchfactor\fontdimen3\font minus
  \fontdimen4\font\relax}
\providecommand{\BIBforeignlanguage}[2]{{%
\expandafter\ifx\csname l@#1\endcsname\relax
\typeout{** WARNING: IEEEtran.bst: No hyphenation pattern has been}%
\typeout{** loaded for the language `#1'. Using the pattern for}%
\typeout{** the default language instead.}%
\else
\language=\csname l@#1\endcsname
\fi
#2}}
\providecommand{\BIBdecl}{\relax}
\BIBdecl

\bibitem{GunduzD2014}
D.~G\"{u}nd\"{u}z, K.~Stamatiou, N.~Michelusi, and M.~Zorzi, ``Designing
  intelligent energy harvesting communication systems,'' \emph{IEEE Commun.
  Mag.}, vol.~52, no.~1, pp. 210--216, Jan. 2014.

\bibitem{HoC2013}
C.~K. Ho, P.~H. Tan, and S.~Sun, ``Energy-efficient relaying over multiple
  slots with causal {CSI},'' \emph{IEEE J. Sel. Areas Commun.}, vol.~31, no.~8,
  pp. 1494--1505, Aug. 2013.

\bibitem{YangJ2012TWC}
J.~Yang, O.~Ozel, and S.~Ulukus, ``Broadcasting with an energy harvesting
  rechargeable transmitter,'' \emph{IEEE Trans. Wireless Commun.}, vol.~11,
  no.~2, pp. 571--583, Feb. 2012.

\bibitem{YangJ2012TC}
J.~Yang and S.~Ulukus, ``Optimal packet scheduling in an energy harvesting
  communication system,'' \emph{IEEE Trans. Commun.}, vol.~60, no.~1, pp.
  220--230, Jan. 2012.

\bibitem{TutuncuogluK2012TWC}
K.~Tutuncuoglu and A.~Yener, ``Optimum transmission policies for battery
  limited energy harvesting nodes,'' \emph{IEEE Trans. Wireless Commun.},
  vol.~11, no.~3, pp. 1180--1189, Mar. 2012.

\bibitem{TutuncuogluK2012JCN}
------, ``Sum-rate optimal power policies for energy harvesting transmitters in
  an interference channel,'' \emph{J. Commun. Netw.}, vol.~14, no.~2, pp.
  151--161, Apr. 2012.

\bibitem{ZaferM2009}
M.~A. Zafer and E.~Modiano, ``A calculus approach to energy-efficient data
  transmission with quality-of-service constraints,'' \emph{IEEE/ACM Trans.
  Netw.}, vol.~17, no.~3, pp. 898--911, June 2009.

\bibitem{SharmaV2010}
V.~Sharma, U.~Mukherji, V.~Joseph, and S.~Gupta, ``Optimal energy management
  policies for energy harvesting sensor nodes,'' \emph{IEEE Trans. Wireless
  Commun.}, vol.~9, no.~4, pp. 1326--1336, Apr. 2010.

\bibitem{SrivastavaR2013}
R.~Srivastava and C.~E. Koksal, ``Basic performance limits and tradeoffs in
  energy-harvesting sensor nodes with finite data and energy storage,''
  \emph{IEEE/ACM Trans. Netw.}, vol.~21, no.~4, pp. 1049--1062, Aug. 2013.

\bibitem{OzelO2011}
O.~Ozel, K.~Tutuncuoglu, J.~Yang, S.~Ulukus, and A.~Yener, ``Transmission with
  energy harvesting nodes in fading wireless channels: Optimal policies,''
  \emph{IEEE J. Sel. Areas Commun.}, vol.~29, no.~8, pp. 1732--1743, Sept.
  2011.

\bibitem{HoC2012}
C.~K. Ho and R.~Zhang, ``Optimal energy allocation for wireless communications
  with energy harvesting constraints,'' \emph{IEEE Trans. Signal Process.},
  vol.~60, no.~9, pp. 4808--4818, Sept. 2012.

\bibitem{ZhangF2014}
F.~Zhang and V.~Lau, ``Closed-form delay-optimal power control for energy
  harvesting wireless system with finite energy storage,'' \emph{IEEE Trans.
  Signal Process.}, vol.~62, no.~21, pp. 5706--5715, Nov. 2014.

\bibitem{MaoZ2012}
Z.~Mao, C.~Koksal, and N.~Shroff, ``Near optimal power and rate control of
  multi-hop sensor networks with energy replenishment: Basic limitations with
  finite energy and data storage,'' \emph{IEEE Trans. Autom. Control}, vol.~57,
  no.~4, pp. 815--829, Apr. 2012.

\bibitem{WangZ2012}
Z.~Wang, A.~Tajer, and X.~Wang, ``Communication of energy harvesting tags,''
  \emph{IEEE Trans. Commun.}, vol.~60, no.~4, pp. 1159--1166, Apr. 2012.

\bibitem{MaoY2014}
Y.~Mao, G.~Yu, and Z.~Zhang, ``On the optimal transmission policy in hybrid
  energy supply wireless communication systems,'' \emph{IEEE Trans. Wireless
  Commun.}, vol.~13, no.~11, pp. 6422--6430, Nov. 2014.

\bibitem{BlascoP2013}
P.~Blasco, D.~Gunduz, and M.~Dohler, ``A learning theoretic approach to energy
  harvesting communication system optimization,'' \emph{IEEE Trans. Wireless
  Commun.}, vol.~12, no.~4, pp. 1872--1882, Apr. 2013.

\bibitem{HuangL2013_TN}
L.~Huang and M.~J. Neely, ``Utility optimal scheduling in energy-harvesting
  networks,'' \emph{IEEE/ACM Trans. Netw.}, vol.~21, no.~4, pp. 1117--1130,
  Aug. 2013.

\bibitem{ZhouK1996_BOOK}
K.~Zhou, J.~C. Doyle, K.~Glover \emph{et~al.}, \emph{Robust and optimal
  control}.\hskip 1em plus 0.5em minus 0.4em\relax Prentice Hall New Jersey,
  1996.

\bibitem{BenTalA2009}
A.~Ben-Tal, L.~El~Ghaoui, and A.~Nemirovski, \emph{Robust optimization}.\hskip
  1em plus 0.5em minus 0.4em\relax Princeton University Press, 2009.

\bibitem{AntunesD2014}
D.~Antunes and W.~Heemels, ``Rollout event-triggered control: Beyond periodic
  control performance,'' \emph{IEEE Trans. Autom. Control}, vol.~59, no.~12,
  pp. 3296--3311, Dec. 2014.

\bibitem{VazeR2014}
R.~Vaze, R.~Garg, and N.~Pathak, ``Dynamic power allocation for maximizing
  throughput in energy-harvesting communication system,'' \emph{IEEE/ACM Trans.
  Netw.}, vol.~22, no.~5, pp. 1621--1630, Oct. 2014.

\bibitem{GomezVilardeboJ2014}
J.~Gomez-Vilardebo and D.~Guenduez, ``Competitive analysis of energy harvesting
  wireless communication systems,'' in \emph{Eur. Wireless (EW) Conf.}, May
  2014, pp. 1--6.

\bibitem{VazeR2013}
R.~Vaze, ``Competitive ratio analysis of online algorithms to minimize packet
  transmission time in energy harvesting communication system,'' in \emph{Proc.
  IEEE INFOCOM}, Apr. 2013, pp. 115--1123.

\bibitem{BorodinA2005BOOK}
A.~Borodin and R.~El-Yaniv, \emph{Online computation and competitive
  analysis}.\hskip 1em plus 0.5em minus 0.4em\relax Cambridge Univ. Press,
  2005.

\bibitem{StollR1979BOOK}
R.~R. Stoll, \emph{Set theory and logic}.\hskip 1em plus 0.5em minus
  0.4em\relax Courier Corporation, 1979.

\bibitem{JohnsonL1987Patent}
L.~Johnson, ``Lithium battery energy monitor,'' United State Patent
  US4\,693\,119 A, Sept., 1987.

\bibitem{DevillersB2012}
B.~Devillers and D.~Gunduz, ``A general framework for the optimization of
  energy harvesting communication systems with battery imperfections,''
  \emph{J. Commun. Netw.}, vol.~14, no.~2, pp. 130--139, Apr. 2012.

\bibitem{CorlessR1996}
R.~Corless, G.~Gonnet, D.~Hare, D.~Jeffrey, and D.~Knuth,
  ``\BIBforeignlanguage{English}{On the lambertw function},''
  \emph{\BIBforeignlanguage{English}{Advances in Computational Mathematics}},
  vol.~5, no.~1, pp. 329--359, 1996.

\bibitem{GuendaL2014}
L.~Guenda, E.~Santana, A.~Collado, K.~Niotaki, N.~B. Carvalho, and
  A.~Georgiadis, ``Electromagnetic energy harvesting—global information
  database,'' \emph{Trans. on Emerging Telecommun. Technol.}, vol.~25, no.~1,
  pp. 56--63, Jan. 2014.

\end{thebibliography}

\end{document}